\documentclass[aps,pra,reprint,groupedaddress,showkeys]{revtex4-1}
\usepackage{amsmath,graphicx,hyperref,verbatimbox,array,float,amsthm}
\usepackage{indentfirst}
\usepackage{physics}
\usepackage{diagbox}
\usepackage{amssymb}
\usepackage{bbold}
\usepackage{colortbl}
\usepackage{multirow}
\newcommand{\figref}[1]{Fig.~\ref{#1}}

\renewcommand{\eqref}[1]{Eq.~\ref{#1}}
\newcommand{\centered}[1]{\begin{tabular}{l} #1 \end{tabular}}
\newtheorem{theorem}{Theorem}

\begin{document}
\title{Entanglement dynamics governed by time-dependent quantum generators}
\author{Artur Czerwinski}\email[]{aczerwin@umk.pl} 
\affiliation{Institute of Physics, Faculty of Physics, Astronomy and Informatics \\ Nicolaus Copernicus University in Torun, ul. Grudziadzka 5, 87--100 Torun, Poland}

\begin{abstract}
In the article, we investigate entanglement dynamics defined by time-dependent linear generators. We consider multilevel quantum systems coupled to an environment that induces decoherence and dissipation, such that the relaxation rates depend on time. By applying the condition of partial commutativity, one can precisely describe the dynamics of selected subsystems. More specifically, we investigate the dynamics of entangled states. The concurrence is used to quantify the amount of two-qubit entanglement in the time domain. The framework appears an efficient tool for investigating quantum evolution of entangled states driven by time-local generators. In particular, non-Markovian effects can be included to observe a restoration of entanglement in time.
\end{abstract}
\maketitle

\section{Introduction}

Dynamics of open quantum systems can be described by differential equations (master equations) that convey information about the interactions between the system and its environment. However, quite often, master equations are not exactly solvable due to their complexity. Therefore, searching for new methods of integrating the evolution equations remains a vital topic within the theory of open quantum systems.

A celebrated master equation describes evolution of open quantum systems governed by a linear operator $\mathbb{L}: \mathbb{M}_d (\mathbb{C}) \rightarrow \mathbb{M}_d (\mathbb{C})$, where we assume that the space is finite-dimensional \cite{Gorini1976,Lindblad1976,Manzano2020}. The linear operator $\mathbb{L}$ is commonly referred to as the GKSL generator. In such a case, the dynamical map is equivalent to a semigroup:
\begin{equation}\label{eq34}
\rho (t) = \mathrm{e} ^{\mathbb{L} t}\, [\rho(0)],
\end{equation}
where $\rho(0)$ denotes the initial density matrix. A master equation with the GKSL generator is the most general type of Markovian and time-homogeneous evolution that preserves the trace and positivity.

A generalized master equation can be obtained if we assume that the linear generator depends on time:
\begin{equation}\label{eq35}
\frac{\dd \rho(t)}{\dd t} = \mathbb{L}(t)\, [\rho(t)],
\end{equation}
where $\mathbb{L}(t)$ is defined on some time interval $\mathcal{I}$. The dynamics \eqref{eq35} can be solved by implementing a time ordering operator, which, in other words, is called "Dyson series" \cite{Dyson1949}. The formal solution does not appear practical for physical applications since we strive to obtain closed-form dynamical maps. Therefore, fundamental problems of the theory of open quantum systems relate to algebraic properties of $\mathbb{L}(t)$ which guarantee that the solution generates a legitimate physical evolution \cite{Alicki2007}.

In particular, we can discuss a time-dependent GKSL generator such that its dissipative part changes in time. More specifically, we investigate $\mathbb{L}(t)$ in the form \cite{Breuer2004}:
\begin{equation}\label{e3.1}
\mathbb{L}(t)\, [\rho] = - i \left[H, \rho \right] + \sum_k \gamma_k (t) \left(V_k \rho V_k^{\dagger} - \frac{1}{2} \left\{V_k ^{\dagger} V_k, \rho \right\}  \right),
\end{equation}
where $V_k^{\dagger}$ stands for the conjugate transpose of $V_k$. This generator involves a physical model, where the jump operators, $V_k$, are represented by constant matrices while the relaxation rates, $\gamma_k (t)$, are time-dependent. The operator $H$ is hermitian, and it can be interpreted as the effective Hamiltonian that accounts for the unitary evolution. The time-local generator \eqref{e3.1} is Hermiticity- and trace-preserving, but for negative relaxation rates it may lead to non-Markovian effects \cite{Breuer2009,Breuer2015}. In this work, we mostly consider positive relaxation rates, which means that the evolution can be called time-dependent Markovian. However, non-Markovianity is also investigated as a separate case.

The master equation of the form \eqref{e3.1} can be implemented for an analysis of quantum systems immersed within an engineered environment \cite{Grigoriu2013}. This approach allows one to address the problem of controllability by the environment (i.e., control by $\gamma_k (t)$), which affects a system through dissipative dynamics and can be used to
steer the system from an initial state (pure or a mixed) towards a designated state \cite{Pechen2006}. Therefore, research on time-dependent quantum generators is strongly motivated by a large number of applications of quantum control, including: quantum computation, quantum engineering, and management of decoherence processes \cite{Chuang1995}.

To solve master equations with generators \eqref{e3.1} we implement the condition of partial commutativity \cite{Kamizawa2018}. This method can be considered a generalization of functional and integral commutativity \cite{Turcotte2002,Kamizawa2015,Maouche2020}. If a generator $\mathbb{L}(t)$ is partially commutative, one can write the closed-form solution for initial density matrices that belong to a subset determined by this condition. The framework for implementation of partial commutativity in dynamics of open quantum systems has already been introduced and applied to specific examples \cite{Czerwinski2020}. The present contribution substantially broadens the scope of the framework by applying it to
investigate the dynamics of entangled states. The model allows one to precisely track different characteristics of entanglement in the time domain.

In this paper, we focus on entangled states, which are a key resource in quantum communication and information \cite{Ekert1991,Horodecki1996,Horodecki2009}. The amount of entanglement can change in time due to the coupling between the system and the environment. For two-qubit states, one can directly compute a measure to quantify entanglement versus time. In particular, one can implement the concurrence \cite{Zhao2011,Ma2012,Menezes2017} or the tangle \cite{Olsen2007}. The analysis of bipartite entanglement may involve two qubits embedded in a common environment \cite{Bratus2021} or two independent baths \cite{ZengZhao2009}.

In Sec.~\ref{review}, we revise the condition of partial commutativity from the point of view of open quantum systems. Then, in Sec.~\ref{twoqubit} and \ref{threequbit}, we implement the framework to investigate the dynamics of two-qubit and three-qubit entangled states, respectively, governed by time-dependent dissipative generators. Next, in Sec.~\ref{twoqutrit}, we study the evolution of two entangled qutrits subject to the same bath. The framework allows one to track how the amount of entanglement declines as the system undergoes a relaxation towards the ground state. Finally, in Sec.~\ref{nonmarkovian}, we consider non-Markovian evolution. The scheme proves to be an efficient tool to observe a backflow of information, for specific time-local quantum generators. Such a phenomenon can lead to the restoration of an entangled state over time.

\section{Partially commutative open quantum systems}\label{review}

For some generators of evolution, the dynamics \eqref{eq35} allows a closed-form solution:
\begin{equation}\label{eq36}
\rho(t) = \exp \left( \int_0^t \mathbb{L}(\tau) d \tau \right) [\rho(0)].
\end{equation}
However, a necessary condition for the generator $\mathbb{L}(t)$ to guarantee a solution in the closed form remains unknown. Up to now, only a few classes of linear differential equations are known to be solvable in closed forms, which justifies further research into the concept of integrability.

In the literature, some sufficient conditions for integrability of the master equation have been determined. In particular, we know that if the generator $\mathbb{L}(t)$ is functionally commutative, one can obtain the closed-form solution, see, e.g., Ref.~\cite{Erugin1966,Goff1981,Zhu1989,Zhu1992}. The class of functionally commutative systems (known also as the Lappo-Danilevsky systems) is well-described, and was also studied in connection to open quantum systems \cite{Kamizawa2015}. A typical subclass of the functionally commutative operators
contains such generators $\mathbb{L}(t)$ that commute with their integrals \cite{Martin1967,Lukes1982}.

In the present article, we investigate the condition of partial commutativity \cite{Fedorov1960,Erugin1966}, which can be considered a generalization of the Lappo-Danilevsky systems. Partial commutativity allows one to follow the closed-form solution for a subset of initial states determined by this criterion. The theorem, which was introduced by Fedorov in 1960, remained unknown for almost 60 years until it was reestablished by Kamizawa in 2018 \cite{Kamizawa2018}. In 2020, it was implemented for quantum dynamics to investigate dissipative multilevel systems with decoherence rates depending on time \cite{Czerwinski2020}. It appears that the applicability of this technique is extensive, which makes it worth studying with connection to evolution of physical systems.

First, the dynamics \eqref{eq35} can always be transformed into a standard matrix equation, where the matrix representation of the generator $\mathbb{L}(t)$ multiplies the vectorized density matrix $\mathrm{vec} [ \rho(t)]$. For any matrix $M$, the operator $\mathrm{vec} [M]$ should be understood as a vector constructed by stacking the columns of $M$ one underneath the other. Thus, let us consider the master equation \eqref{eq35} in the vectorized form, i.e.:
\begin{equation}\label{eq41a}
\mathrm{vec}[\dot{\rho}(t)] = \mathbb{L}(t) \; \mathrm{vec}[ \rho(t) ].
\end{equation}
Particularly, the generator \eqref{e3.1} can be represented as a matrix by following the Roth's column lemma \cite{Roth1934,Henderson1981}. For any three matrices $A, B, C$ (such that product $ABC$ is computable), we can prove :
\begin{equation}\label{e3.2}
\mathrm{vec} \:[ ABC ] = \left(C^T \otimes A\right) \:\mathrm{vec} [B],
\end{equation}
By implementing the Roth's column lemma, one transforms the generator \eqref{e3.1} into its matrix form:
\begin{equation}\label{e3.3}
\begin{aligned}
&\mathbb{L}(t) = i \left( H^T  \otimes \mathbb{1}_d - \mathbb{1}_d \otimes H  \right) \\& + \sum_{k} \gamma_k  (t)  \left ( \overline{V}_k \otimes V_k - \frac{1}{2} \mathbb{1}_d \otimes V_k ^{\dagger} V_k - \frac{1}{2} V_k ^T \overline{V}_k \otimes \mathbb{1}_d \right ),
\end{aligned}
\end{equation}
where $\overline{V}_k$ denotes the complex conjugate of the jump operator $V_k$.

Then, we can formulate the condition of partial commutativity, which is alternatively called the Fedorov theorem \cite{Fedorov1960}.
\begin{theorem}[Fedorov theorem]
If the matrix representation of the generator $\mathbb{L}(t)$ satisfies the condition:
\begin{equation}\label{eq41}
[\,\mathbb{L}(t), B^n (t)\,]\: \alpha = 0 \hspace{0.5cm} \forall \:n=1,2, 3, \dots \hspace{0.35cm}\text{and} \hspace{0.35cm}\forall t\in\mathcal{I},
\end{equation}
where $B(t) = \int_0^t \mathbb{L}(\tau) d \tau$ and $\alpha$ is a constant vector, then there exists a closed-form solution of \eqref{eq41a}:
\begin{equation}\label{eq42}
\mathrm{vec} [ \rho(t)] = \mathrm{e} ^{B(t)} \,\alpha.
\end{equation}
\end{theorem}

The proof of the Fedorov theorem can be found in Ref.~\cite{Czerwinski2020}. The major limitation of the Fedorov theorem concerns the fact that the closed-form solution is admissible only for vectors $\alpha$ that satisfy the condition \eqref{eq41}. This means that we need to determine the subspace of all allowable initial vectors:
\begin{equation}\label{eq47}
\mathcal{M} :=  \bigcap_{t\in\mathcal{I}} \bigcap_{n=1}^{\mu-1} \mathrm{Ker} \left[\,\mathbb{L}(t), B^n (t)\,\right],
\end{equation}
where $\mu$ stands for the degree of the minimal polynomial of $B(t)$ (i.e., the matrix polynomial of the lowest degree, such that $B(t)$ is a root of the polynomial). In the definition of $\mathcal{M}$ we treat $t$ as an independent parameter, which means that the result should be fixed. This allows us to obtain a solution that holds for all $t\in\mathcal{I}$.

The formula \eqref{eq47} cannot be easily calculated. However, one can use the approach introduced by Shemesh to transform this expression into a form that can be computed straightforwardly \cite{Shemesh1984,Jamiolkowski2014}:
\begin{equation}\label{eq49}
\mathcal{M} = \bigcap_{t\in\mathcal{I}} \mathrm{Ker} \sum_{n=1}^{\mu-1}\, [\,\mathbb{L}(t), B^n(t)\,]^{\dagger} [\,\mathbb{L}(t), B^n(t)\,].
\end{equation}

To sum up, if one wants to apply the Fedorov theorem to obtain a closed-form solution of a differential equation with a time-dependent generator $\mathbb{L}(t)$, one needs to prove that the subspace $\mathcal{M}$ defined by \eqref{eq47} is non-empty, which can be done effectively by implementing the Shemesh criterion \eqref{eq49}. Then, one gets the closed-form solution according to \eqref{eq42}. The solution defines a legitimate trajectory in the state set if $\alpha$ can be considered a vectorized density matrix, i.e. $\alpha \equiv \mathrm{vec} [\rho(0) ]$. In other words, we operate only within the physically admissible subset of initial vectors: $\alpha \in \mathcal{M} \cap \mathrm{vec} \,[S(\mathcal{H})]$. Time-dependent generators $\mathbb{L}(t)$ that correspond to a non-empty subspace $\mathcal{M}$ can be called \textit{partially commutative}.

\section{Two-qubit entangled states}\label{twoqubit}

We consider cascade systems with three energy levels described by quantum states: $\ket{1}, \ket{2}, \ket{3}$ \cite{Hioe1982,Rooijakkers1997}. Therefore, we operate in the $3-$dimensional Hilbert space and, for simplicity, we assume that the vectors $\{\ket{1}, \ket{2}, \ket{3}\}$ constitute the standard basis in $\mathcal{H}$. Two types of transition are possible: $\ket{3}\rightarrow\ket{2}$ and $\ket{2}\rightarrow\ket{1}$. In other words, the model describes a relaxation towards the ground state $\ket{1}$. Let us assume that this process is governed by a time-local generator \cite{Czerwinski2020}:

\begin{equation}\label{tq1}
\begin{aligned}
\mathbb{L} (t) {}&=  i \left( H^T  \otimes \mathbb{1}_3 - \mathbb{1}_3 \otimes H \right)  \\ & +\mathrm{sin}^2\omega t \left (E_{23} \otimes E_{23} - \frac{1}{2} \mathbb{1}_3 \otimes E_{33} -\frac{1}{2} E_{33} \otimes \mathbb{1}_3  \right) \\
& + \mathrm{cos}^2 \omega t \left(E_{12} \otimes E_{12} - \frac{1}{2} \mathbb{1}_3 \otimes E_{22} - \frac{1}{2} E_{22} \otimes \mathbb{1}_3  \right),
\end{aligned}
\end{equation}
where $H$ denotes the unperturbed Hamiltonian with three symmetric energy levels, i.e. $H = \mathrm{diag} (-\mathcal{E}, 0, \mathcal{E} )$ and $E_{ij} = \ket{i} \!\bra{j}$ represents the jump operator. Additionally, $\omega$ stands for the angular frequency characterizing the dynamics.

The dynamics governed by \eqref{tq1} has a closed-form solution for such initial states that have zero probability of occupying the highest energy level \cite{Czerwinski2020}. Thus, the condition of partial commutativity allows one to precisely describe the evolution of two-level systems immersed within the $3-$dimensional Hilbert space. This fact gives the gist of the Fedorov theorem -- dynamical maps in the closed forms are obtainable only for a restricted subset of states.

Furthermore, if we have a pair of two-level systems (denoted by $A$ and $B$) and each of them is subject to \eqref{tq1}, we can describe the dynamics by a joint two-qubit generator:
\begin{equation}\label{2qgen}
\mathbb{L}_{2q} (t) = \mathbb{L}^{(A)} (t) \otimes \mathbb{I}^{(B)}_9 +  \mathbb{I}^{(A)}_9 \otimes\mathbb{L}^{(B)} (t),
\end{equation}
which is known as the Kronecker sum: $\mathbb{L}_{2q} (t) \equiv \mathbb{L}^{(A)} (t) \oplus \mathbb{L}^{(B)} (t)$. For any initial state $\rho^{AB} (0)$ such that neither of the subsystems can be found in the highest energy level, we can follow the closed-form trajectory
\begin{equation}\label{2qmap}
\rho^{AB} (t) = \exp \left( \int_0^t \mathbb{L}_{2q} (\tau) d \tau \right) [\rho^{AB} (0)].
\end{equation}
Therefore, starting from three-level dynamics \eqref{tq1}, we can describe the dynamics of two-qubit entangled states within the framework of partial commutativity.

\subsection{Example 1: Evolution of $\ket{\Phi (\phi)}$}

Let us consider the dynamics of bipartite entanglement subject to the generator \eqref{2qgen} with the initial state given as $\rho^{AB} (0) =\ket{\Phi (\phi)}\!\bra{\Phi (\phi)}$, where
\begin{equation}\label{ex1}
\ket{\Phi (\phi)} = \frac{1}{\sqrt{2}} \left( \ket{1}_A \otimes \ket{1}_B + e ^{i \phi} \ket{2}_A \otimes \ket{2}_B \right)
\end{equation}
with $0\leq \phi < 2 \pi$ standing for the relative phase. This class of entanglement includes the two celebrated Bell states: $\ket{\Phi^+}$ and $\ket{\Phi^-}$. Such an initial state satisfies the condition of partial commutativity since \eqref{ex1} is a superposition of the middle and the ground state. This implies that the dynamical map can be computed in the closed-form based on \eqref{2qmap}. We neglect the elements of the density matrix that relate to the highest energy level since it cannot be occupied. Then, by implementing a mathematical software to solve \eqref{2qmap}, one obtains a $4\times4$ density matrix that describes the dynamics of two-qubit entanglement:
\begin{widetext}
\begin{equation}\label{ex2}
\rho^{AB} (t)= \frac{1}{2} \begin{pmatrix} 1 + (1-\xi(t))^2 & 0 & 0&  \xi(t) \: e^{ i (\phi + 2 \mathcal{E} t) } \\ 0 & \xi(t) \left(1 - \xi(t) \right) & 0& 0 \\ 0 & 0 &\xi(t) \left(1 - \xi(t) \right)& 0 \\  \xi(t) \:e^{ - i (\phi + 2 \mathcal{E} t) } & 0 & 0& \xi^2(t) \end{pmatrix}
\end{equation}
\end{widetext}
where $\xi(t) = e ^{- \frac{2 \omega t + \sin 2 \omega t}{4 \omega}}$.

\begin{figure}[h!]
\centering
		\centered{\includegraphics[width=0.9\columnwidth]{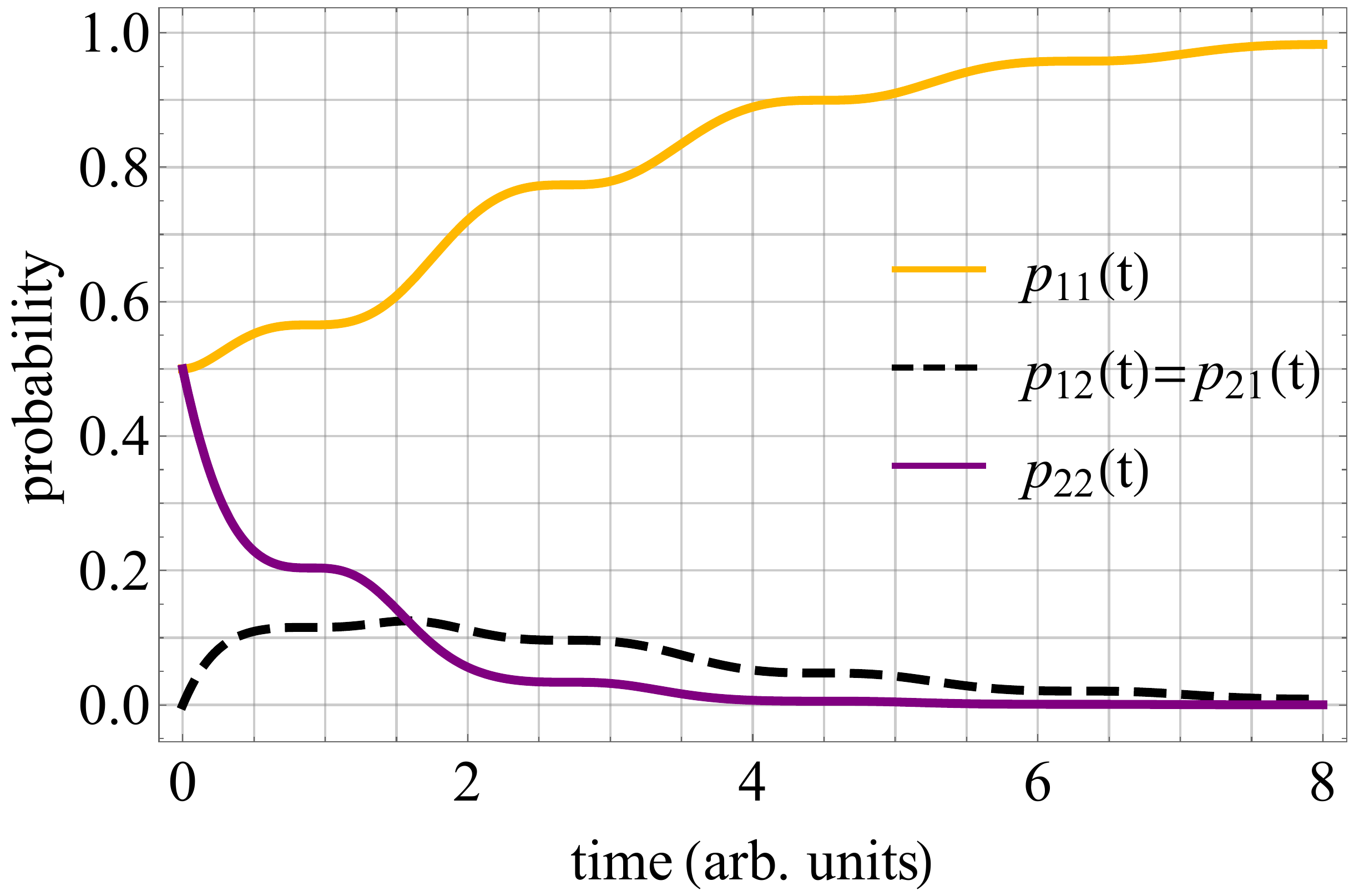}}
	\caption{Plots present the probability of finding a two-qubit system in one of the possible states. The initial state is represented by \eqref{ex1}.}
	\label{figure1}
\end{figure}

First, by $p_{ij} (t)$, we denote the joint probability of finding upon measurement the subsystems $A$ and $B$ in the states described by the vectors $\ket{i}$ and $\ket{j}$, respectively. Then, one can track the probabilities versus time, which is presented in \figref{figure1} for an arbitrary $\omega$. We observe that $p_{11} (t)$ gradually increases whereas $p_{22} (t)$ declines, which reflects the fact that the dynamics describe the relaxation towards the ground level. However, for $t>0$, we notice non-zero values of $p_{12} (t)$ (and $p_{21} (t)$), which implies that the states are not perfectly correlated. It may happen that the $A-$system has already decayed to the ground level, but the $B-$system remains in the middle level (or vice versa).

\begin{figure}[h!]
\centering
		\centered{\includegraphics[width=0.9\columnwidth]{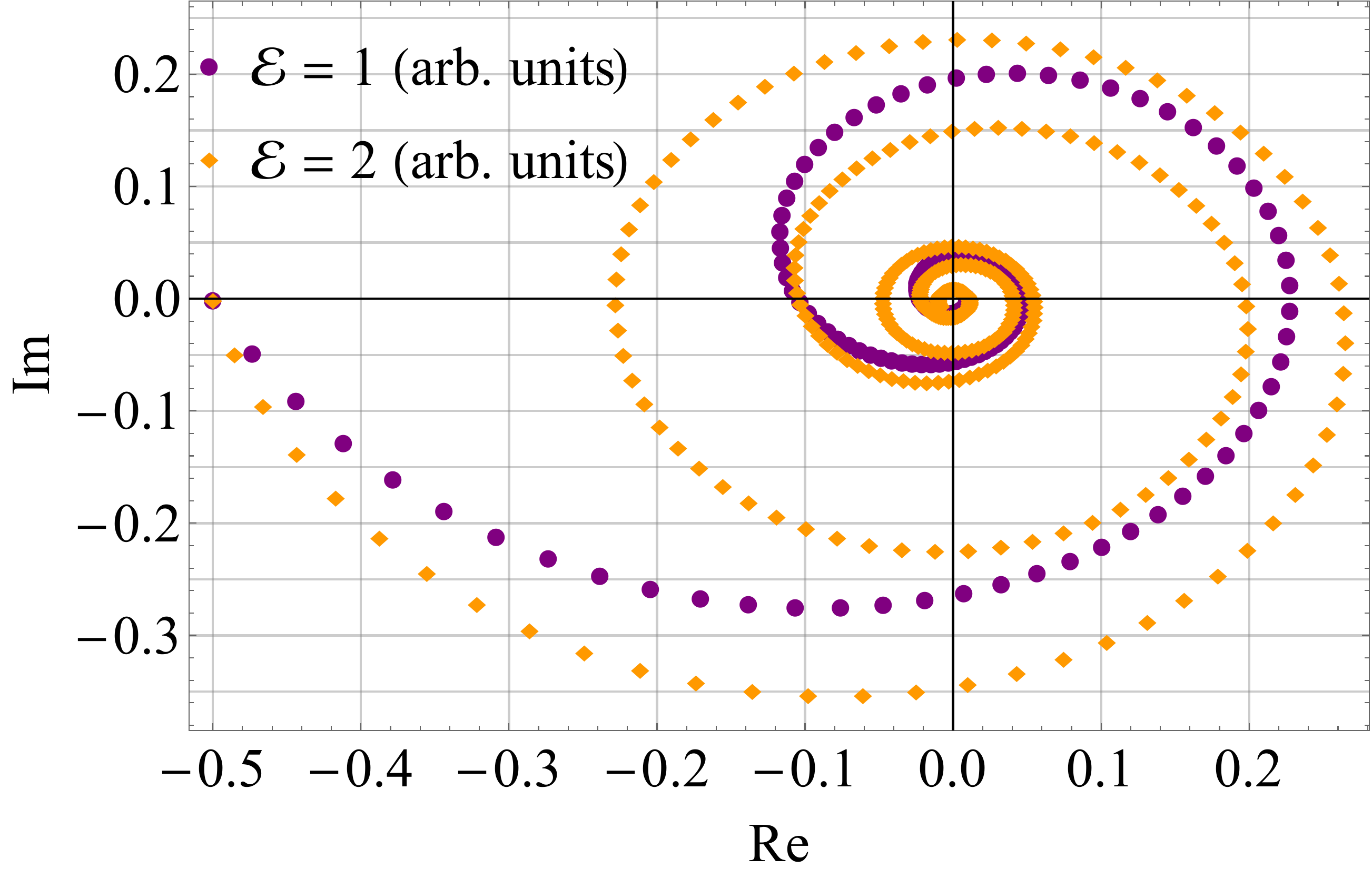}}
	\caption{Trajectories of the phase factor, $\rho^{AB}_{14} (t)$, presented on the complex plane for two values of $\mathcal{E}$.}
	\label{figure2}
\end{figure}

The diagonal elements of the density matrix \eqref{ex2} are not affected by $\mathcal{E}$ or $\phi$. The energy levels of the unperturbed Hamiltonian govern the evolution of the phase factor on the complex plane. Assuming $\phi = \pi$ and $\omega$ is fixed arbitrarily, one can follow the trajectories of the off-diagonal elements of the density matrix $\rho^{AB} (t)$. In \figref{figure2}, two trajectories are presented. One can agree that the dynamics of $\rho^{AB}_{14} (t)$ features two aspects. First, the phase factor rotates on the complex plane, which is caused by $\mathcal{E}$. Secondly, $\rho^{AB}_{14} (t)$ approaches zero as time grows, which can be interpreted as a phase-damping effect brought about by the dissipative part of the generator of evolution.

\begin{figure}[h!]
\centering
		\centered{\includegraphics[width=0.9\columnwidth]{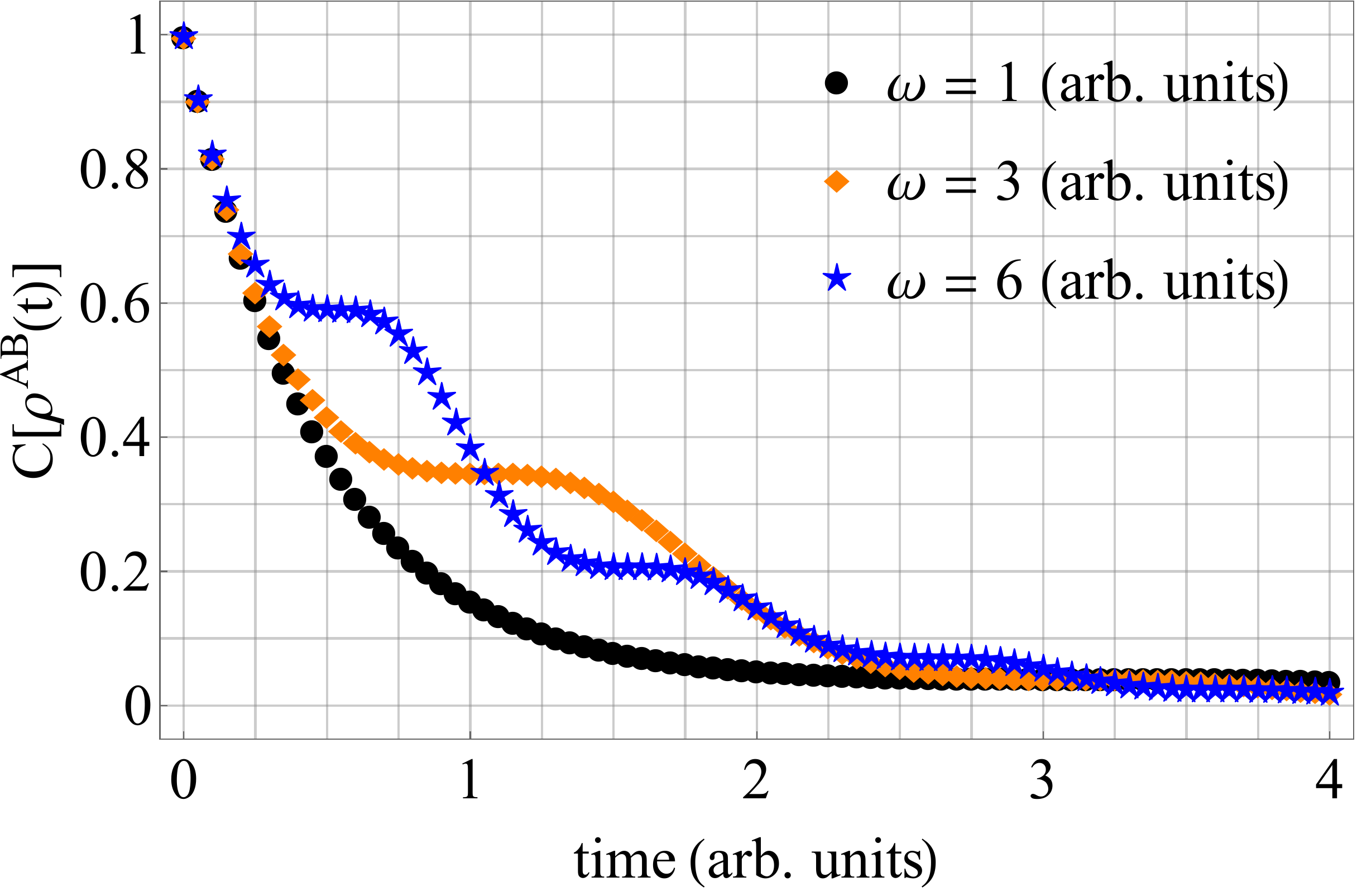}}
	\caption{Concurrence, $C[\rho^{AB} (t)]$, of the two-qubit density matrix with the initial state \eqref{ex1} for three values of $\omega$.}
	\label{figure3}
\end{figure}

Furthermore, we investigate how the amount of entanglement changes in time. Thus, the concurrence \cite{Hill1997,Wootters1998}, denoted by $C[\rho^{AB} (t)]$, is computed and presented in \figref{figure3}. The properties of $C[\rho^{AB} (t)]$ are influenced by $\omega$ (and not by $\mathcal{E}$). For this reason, three plots corresponding to different values of $\omega$ are shown. For every plot, we have $C[\rho^{AB} (0)] = 1$ since the initial state is maximally mixed (irrespective of $\phi$). One can notice that all plots are non-increasing and converge to zero with time, which stems from the properties of the generator of evolution. However, the pace of entanglement decline is different. Based on the plots, one can predict how much entanglement is preserved after a given period of time.

\subsection{Example 2: Evolution of $\ket{\Psi (\phi)}$}

Let us consider another class of maximally entangled two-qubit states:
\begin{equation}\label{ex3}
\rho^{AB} (0) =\ket{\Psi (\phi)}\!\bra{\Psi (\phi)},
\end{equation}
where
\begin{equation}\label{ex4}
\ket{\Psi (\phi)} = \frac{1}{\sqrt{2}} \left( \ket{1}_A \otimes \ket{2}_B + e ^{i \phi} \ket{2}_A \otimes \ket{1}_B \right),
\end{equation}
which includes the other two famous Bell states: $\ket{\Psi^+}$ and $\ket{\Psi^-}$. For any relative phase, the state \eqref{ex4} represents perfectly anti-correlated results, which means that if the subsystem $A$ is measured to be in the state $\ket{1}$, the subsystem $B$ is bound to be in $\ket{2}$ (and vice versa).

For input states of the form \eqref{ex3}, we obtain a closed-form solution according to \eqref{2qmap}. Since the highest energy level is forbidden, we can reduce the output density matrix by eliminating the elements related to $\ket{3}$. Thus, in the same vein as in the above example, we obtain a $4\times4$ density matrix that describes two-qubit entanglement immersed in a higher-dimensional Hilbert space: 
\begin{widetext}
\begin{equation}\label{ex5}
\rho^{AB} (t) = \frac{1}{2} \begin{pmatrix} 2\left(1-e^{-\frac{2 \omega t + \sin 2 \omega t}{4 \omega}}\right) & 0 &0&0\\0&  e ^{- \frac{2 \omega t + \sin 2 \omega t}{4 \omega}} &  e^{-\frac{2(t+ 2 \phi i) \omega + \sin 2 \omega t}{4 \omega}}&0 \\ 0 &  e^{-\frac{2(t- 2 \phi i) \omega + \sin 2 \omega t}{4 \omega}} & e^{- \frac{2 \omega t + \sin 2 \omega t}{4 \omega}}&0 \\0 & 0 & 0& 0  \end{pmatrix}.
\end{equation}
\end{widetext}

\begin{figure}[h!]
\centering
		\centered{\includegraphics[width=0.9\columnwidth]{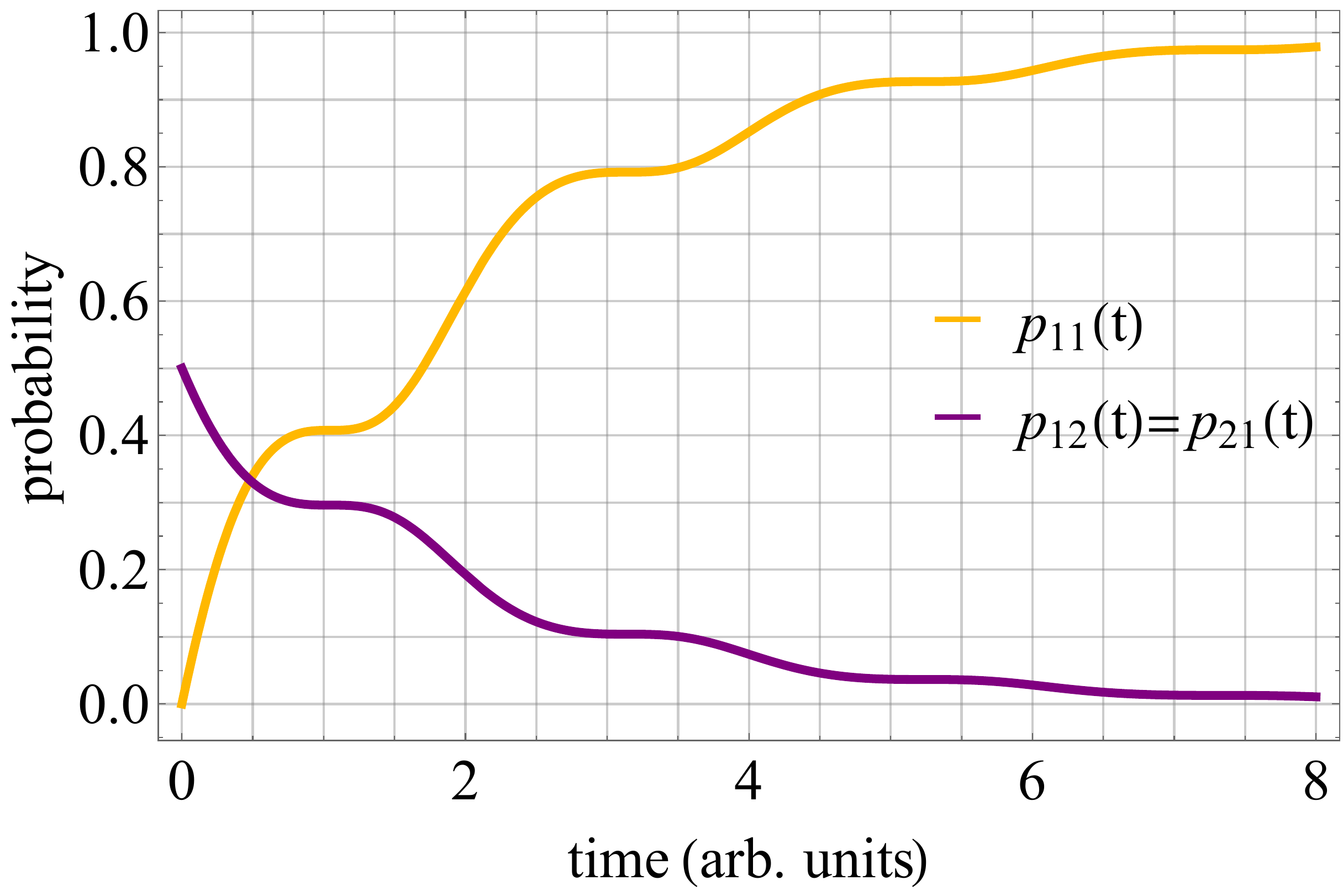}}
	\caption{Plots present the probability of finding a two-qubit system in one of the possible states. The initial state is represented by \eqref{ex4}.}
	\label{figure4}
\end{figure}

First, one can notice that $p_{22} (t) = 0$, which come as a natural consequence of the initial quantum state \eqref{ex4}. Thus, only three configurations of the system are possible. For an arbitrary $\omega$, the probabilities corresponding to the admissible configurations are presented in \figref{figure4}. One can observe that the probability of the anti-correlated configurations declines while $p_{11} (t)$ increases.

\begin{figure}[h!]
\centering
		\centered{\includegraphics[width=0.9\columnwidth]{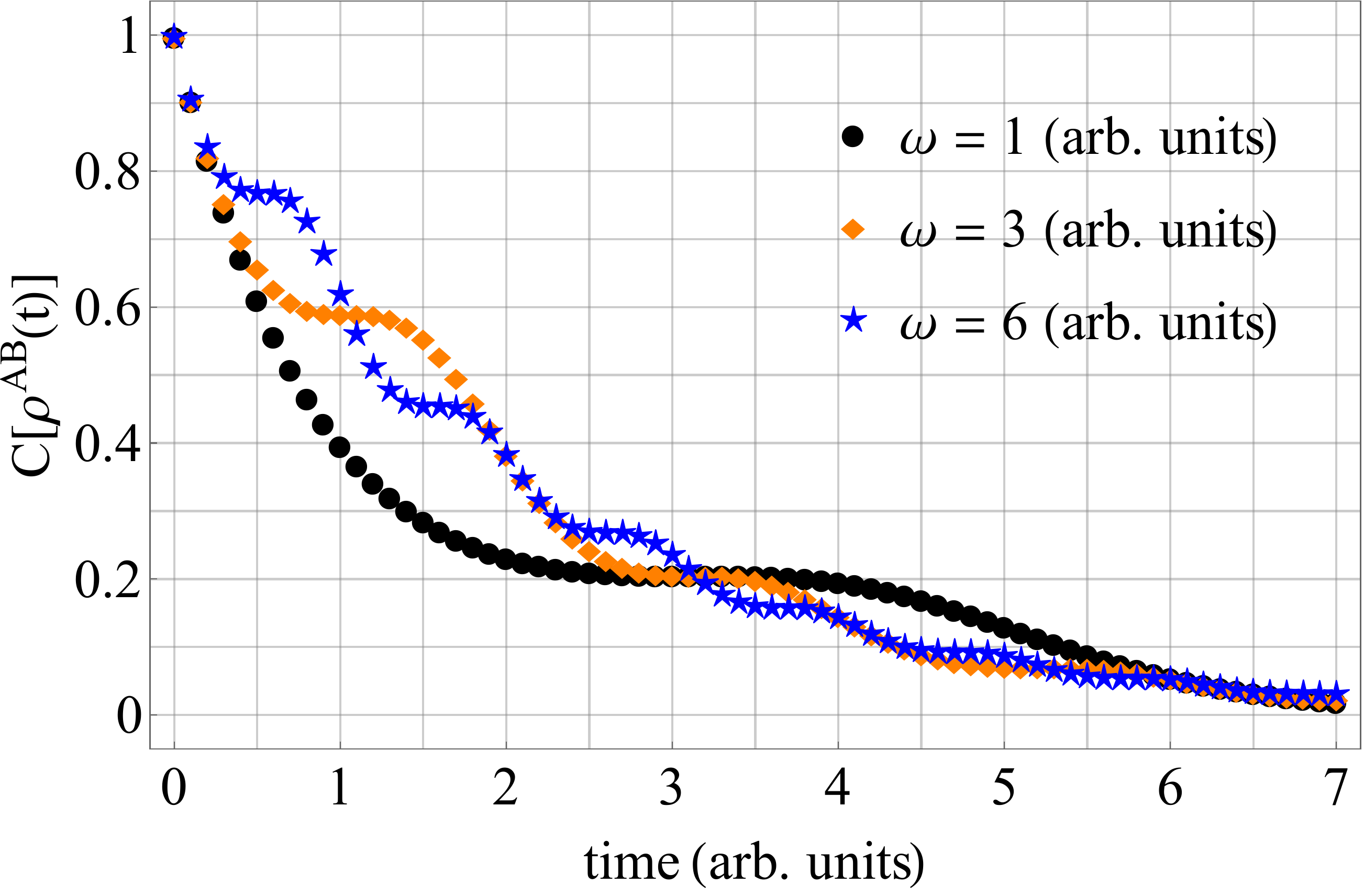}}
	\caption{Concurrence, $C[\rho^{AB} (t)]$, of the two-qubit density matrix with the initial state \eqref{ex4} for three values of $\omega$.}
	\label{figure5}
\end{figure}
Furthermore, the concurrence is investigated as a function of time. For three values of $\omega$, the concurrence is depicted in \figref{figure5}. The plots interlace with one another. 

\section{Three-qubit entangled states}\label{threequbit}

The framework based on partial commutativity can also be applied to three-qubit entangled states. Let us assume that there is a tripartite system with respective components denoted by $A$, $B$, and $C$. We introduce a time-dependent three-qubit generator of evolution:
\begin{equation}\label{ex6}
\begin{aligned}
\mathbb{L}_{3q} (t) = {}&\mathbb{L}^{(A)} (t) \otimes \mathbb{I}^{(B)}_9 \otimes \mathbb{I}^{(C)}_9 +  \mathbb{I}^{(A)}_9 \otimes\mathbb{L}^{(B)} (t)\otimes \mathbb{I}^{(C)}_9\\& +\mathbb{I}^{(A)}_9 \otimes \mathbb{I}^{(B)}_9 \otimes\mathbb{L}^{(C)} (t),
\end{aligned}
\end{equation}
where $\mathbb{L}^{(i)} (t)$ stands for a generator of the form \eqref{tq1}. Just as before, the generator $\mathbb{L}^{(i)} (t)$ describes the evolution of a three-level system. However, to make to master equation solvable, we cannot admit the highest energy level. Therefore, each subsystem allows for a realization of a qubit state. For input states satisfying this condition, $\rho^{ABC} (0)$, one can follow the closed-form solution
\begin{equation}\label{ex7}
\rho^{ABC} (t) = \exp \left( \int_0^t \mathbb{L}_{3q} (\tau) d \tau \right) [\rho^{ABC} (0)].
\end{equation}

\subsection{Example 1: Evolution of the GHZ state}

In particular, we can consider an initial state: $\rho^{ABC} (0) = \ket{GHZ} \! \bra{GHZ}$, where
\begin{equation}\label{ex8}
\ket{GHZ}  = \frac{1}{\sqrt{2}} \left( \ket{1}_A \otimes \ket{1}_B \otimes \ket{1}_C + e ^{ i \phi} \ket{2}_A \otimes \ket{2}_B \otimes \ket{2}_C \right).
\end{equation}
and $0 \leq \phi < 2 \pi$. For $\phi = 0$, the vector \eqref{ex8} represents the Greenberger–Horne–Zeilinger state (GHZ state) \cite{Greenberger1990}, which is celebrated for its importance in quantum information, including quantum teleportation \cite{Karlsson1998}, quantum secret sharing \cite{Hillery1999}, or quantum cryptography \cite{Kempe1999}. By applying the dynamical map \eqref{ex7}, one can track the evolution of the GHZ state with an arbitrary phase. Having reduced the density matrix by eradicating the forbidden level, one obtains a $8\times8$ density matrix that describes the three-qubit state:
\begin{widetext}
\begin{equation}\label{ex9}
\begin{aligned}
{}&\rho^{ABC} (t)=\\& \frac{1}{2} \begin{pmatrix} 1+ \left(1 - e^{-\frac{2 \omega t + \sin 2 \omega t}{4 \omega}}  \right)^3& 0 &0&0&0&0&0& e^{(- \frac{3}{4} + 3 \mathcal{E} i )t - i \phi - \frac{3 \sin 2 \omega t}{8 \omega}}\\0&\Xi (t) &0&0&0&0&0&0\\0&0&\Xi (t)&0&0&0&0&0\\0&0&0&\Sigma (t)&0&0&0&0\\0&0&0&0&\Xi (t)&0&0&0\\0&0&0&0&0&\Sigma (t)&0&0\\0&0&0&0&0&0&\Sigma (t)&0\\e^{(- \frac{3}{4} + 3 \mathcal{E} i )t + i \phi - \frac{3 \sin 2 \omega t}{8 \omega}}&0&0&0&0&0&0& e^{-\frac{3 t}{2} - \frac{3 \sin 2 \omega t}{4 \omega}}  \end{pmatrix}.
\end{aligned}
\end{equation}
\end{widetext}
where
\begin{equation}
\begin{split}
&\Xi (t) = e^{-\frac{2 \omega t + \sin 2 \omega t}{4 \omega}} \left( 1 - e^{-\frac{2 \omega t + \sin 2 \omega t}{4 \omega}}\right)^2,\\
&\Sigma (t) = e^{-\frac{2 \omega t + \sin 2 \omega t}{2 \omega}} \left( 1 - e^{-\frac{2 \omega t + \sin 2 \omega t}{4 \omega}}\right).
\end{split}
\end{equation}

\begin{figure}[h!]
\centering
		\centered{\includegraphics[width=0.9\columnwidth]{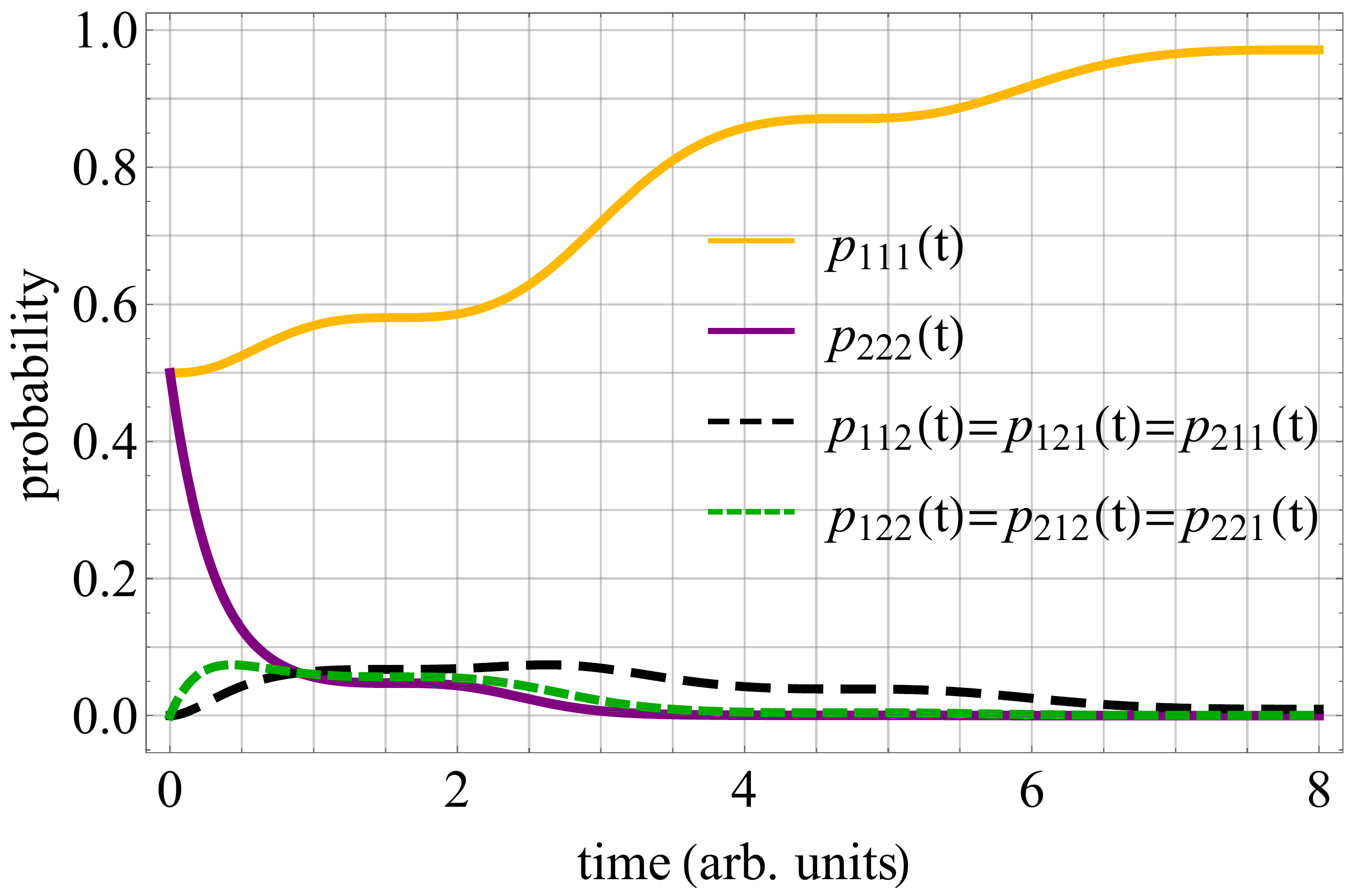}}
	\caption{Plots present the probability of finding a three-qubit system in one of the possible states. The initial state is represented by \eqref{ex8}.}
	\label{figure6}
\end{figure}

In \figref{figure6}, one finds the probabilities of finding the system in one of the possible configurations. We observe that the probability corresponding to $\ket{1}_A \otimes \ket{1}_B \otimes \ket{1}_C$ grows whereas the probability for $\ket{2}_A \otimes \ket{2}_B \otimes \ket{2}_C$ declines, which as an expected tendency. Interestingly, non-zero probabilities relate to other possible configurations. Although all three subsystems are subject to the same bath, it may happen that only one party has already relaxed to the ground state and the other two have not, or two subsystems have collapsed, and one remains in the middle energy level.

\begin{figure}[h!]
\centering
		\centered{\includegraphics[width=0.9\columnwidth]{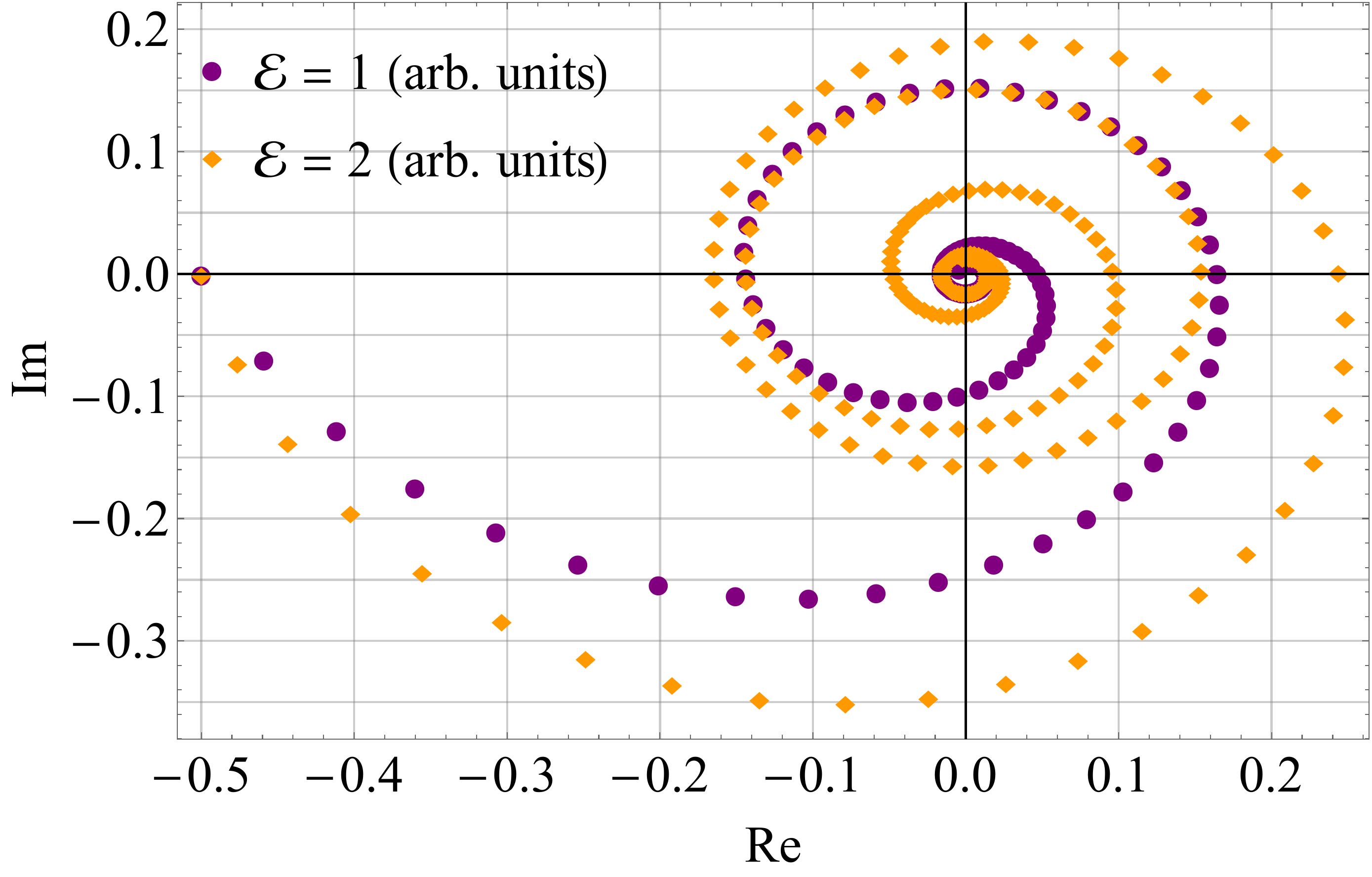}}
	\caption{Trajectories of the phase factor, $\rho^{ABC}_{18} (t)$, presented on the complex plane for two values of $\mathcal{E}$.}
	\label{figure7}
\end{figure}

Then, one can also study the time evolution of the phase factor on the complex plane, which is primarily governed by $\mathcal{E}$. In \figref{figure7}, one finds two trajectories of  $\rho^{ABC}_{18} (t)$ for a fixed value of $\omega$. In this case, we can observe similar tendencies as presented in \figref{figure2}.

\begin{figure}[h!]
\centering
		\centered{\includegraphics[width=0.9\columnwidth]{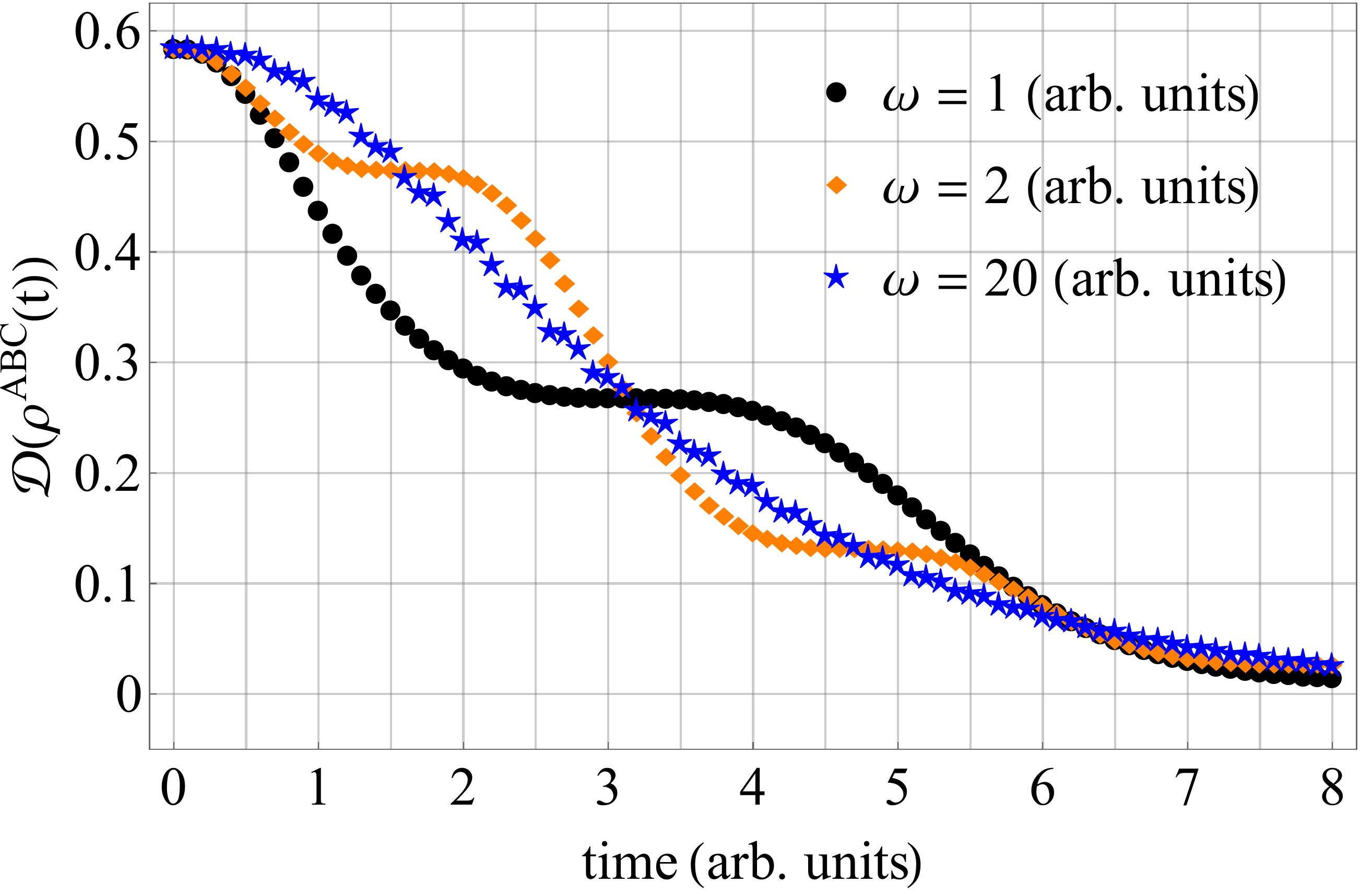}}
	\caption{Plots present the Bures distance, $\mathcal{D} (\rho^{ABC} (t))$, for three values of $\omega$.  Initially, the system was represented by the GHZ state.}
	\label{figure8}
\end{figure}

Furthermore, to describe the decline in entanglement as the initial state evolves, we compute the Bures distance that is given by
\begin{equation}\label{ex10}
\mathcal{D} (\rho^{ABC} (t)) = 2 \left(1 - \sqrt{\mathcal{F}(\rho^{ABC} (t), \rho_{111})}\right),
\end{equation}
where $\rho_{111} = \ket{111}\!\bra{111}$ and, for any two quantum states $\rho$ and $\sigma$, $\mathcal{F}(\rho,\sigma)$ denotes the quantum fidelity defined as \cite{Uhlmann1986,Jozsa1994}
\begin{equation}\label{ex11}
\mathcal{F}(\rho,\sigma) = \left[\tr \sqrt{\sqrt{\rho} \,\sigma\, \sqrt{\rho}}  \right]^2.
\end{equation}
The Bures distance can be implemented to define a geometric measure of entanglement \cite{Vedral1998}. In our application, we consider it a simplified figure to quantify the amount of entanglement. Since the initial state converges to $\ket{111}$ in time, it appears justified to interpret the distance between an instantaneous state $\rho^{ABC} (t)$ and the final state $\ket{111}$ as a measure of entanglement. In \figref{figure8}, one can find three plots obtained for different values of $\omega$. This approach allows us to investigate entanglement dynamics in terms of geometric approaching to the final (separable) state, which reflects the amount of entanglement preserved in the system at a given time.

\subsection{Example 2: Evolution of the W state}

Next, we consider an input state $\rho^{ABC} (t) = \ket{W}\!\bra{W}$, where
\begin{equation}\label{ex12}
\ket{W} = \frac{1}{\sqrt{3}} \left ( \ket{1}_A \ket{1}_B \ket{2}_C  + \ket{1}_A\ket{2}_B\ket{1}_C  + \ket{2}_A\ket{1}_B\ket{1}_C  \right),
\end{equation}
which is commonly referred to as the W state. The states $\ket{GHZ}$ and $\ket{W}$ represent two very different kinds of tripartite entanglement that cannot be transformed into each other by local operations \cite{Dur2000a}. The W state was proposed as a resource for several applications, including secure quantum communication \cite{Jian2007}.

By applying the dynamical map \eqref{ex7}, one obtains the evolution of the W state:
\begin{widetext}
\begin{equation}\label{ex13}
\begin{aligned}
{}& \rho^{ABC} (t) =\\& \frac{1}{3} \begin{pmatrix} 3 \left(1 - e^{-\frac{2 \omega t + \sin 2 \omega t}{4 \omega}}  \right)& 0 &0&0&0&0&0&0\\ 0& e^{-\frac{2 \omega t + \sin 2 \omega t}{4 \omega}} & e^{-\frac{2 \omega t + \sin 2 \omega t}{4 \omega}}&0& e^{-\frac{2 \omega t + \sin 2 \omega t}{4 \omega}}&0&0&0\\ 0& e^{-\frac{2 \omega t + \sin 2 \omega t}{4 \omega}} & e^{-\frac{2 \omega t + \sin 2 \omega t}{4 \omega}}&0& e^{-\frac{2 \omega t + \sin 2 \omega t}{4 \omega}}&0&0&0\\0&0 &0&0&0&0&0&0\\0& e^{-\frac{2 \omega t + \sin 2 \omega t}{4 \omega}} & e^{-\frac{2 \omega t + \sin 2 \omega t}{4 \omega}}&0& e^{-\frac{2 \omega t + \sin 2 \omega t}{4 \omega}}&0&0&0\\0&0 &0&0&0&0&0&0\\0&0 &0&0&0&0&0&0\\0&0 &0&0&0&0&0&0 \end{pmatrix}.
\end{aligned}
\end{equation}
\end{widetext}

\begin{figure}[h!]
\centering
		\centered{\includegraphics[width=0.9\columnwidth]{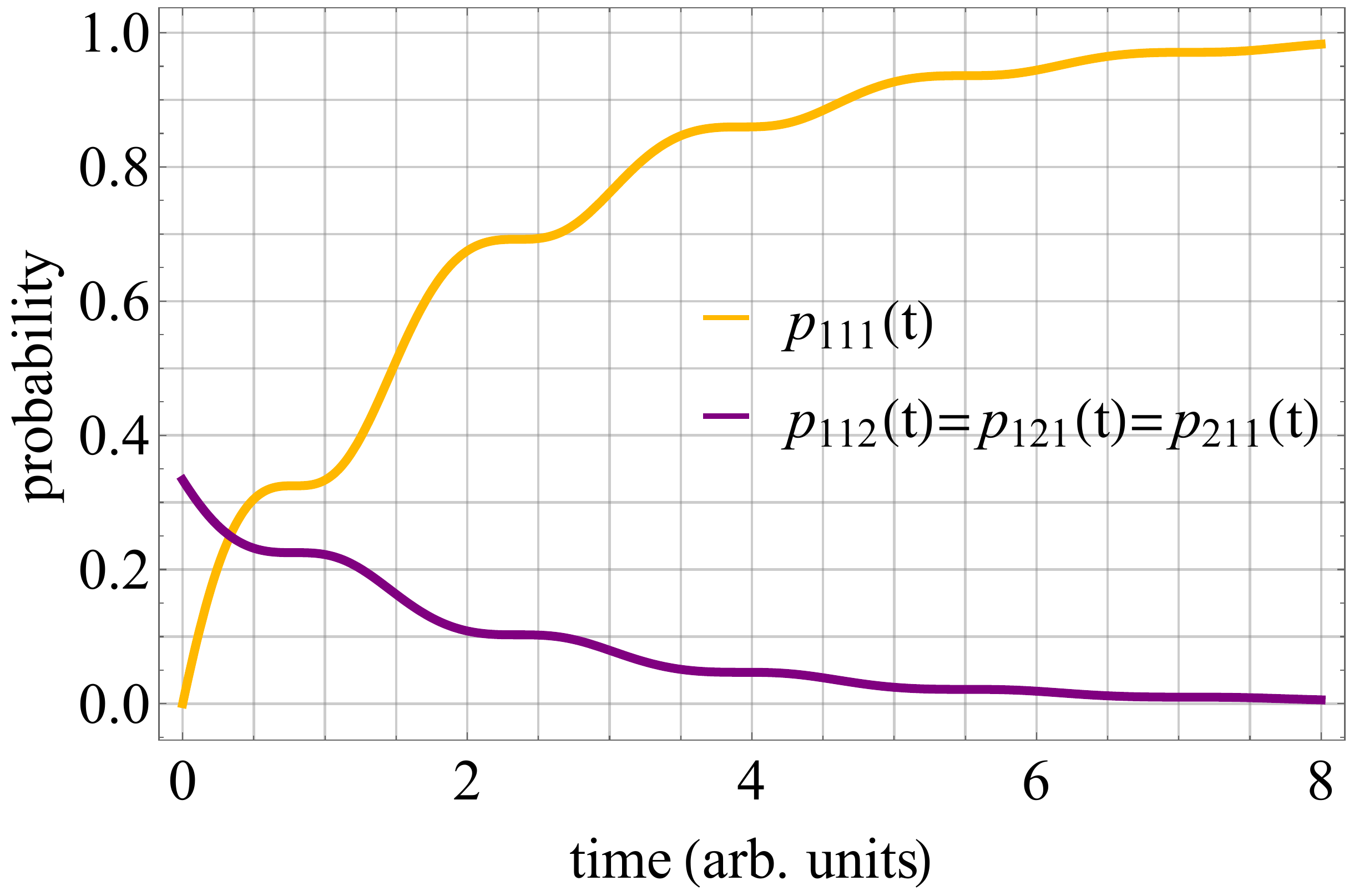}}
	\caption{Plots present the probability of finding a three-qubit system in one of the possible states. The initial state is represented by \eqref{ex12}.}
	\label{figure9}
\end{figure}

\begin{figure}[h!]
\centering
		\centered{\includegraphics[width=0.9\columnwidth]{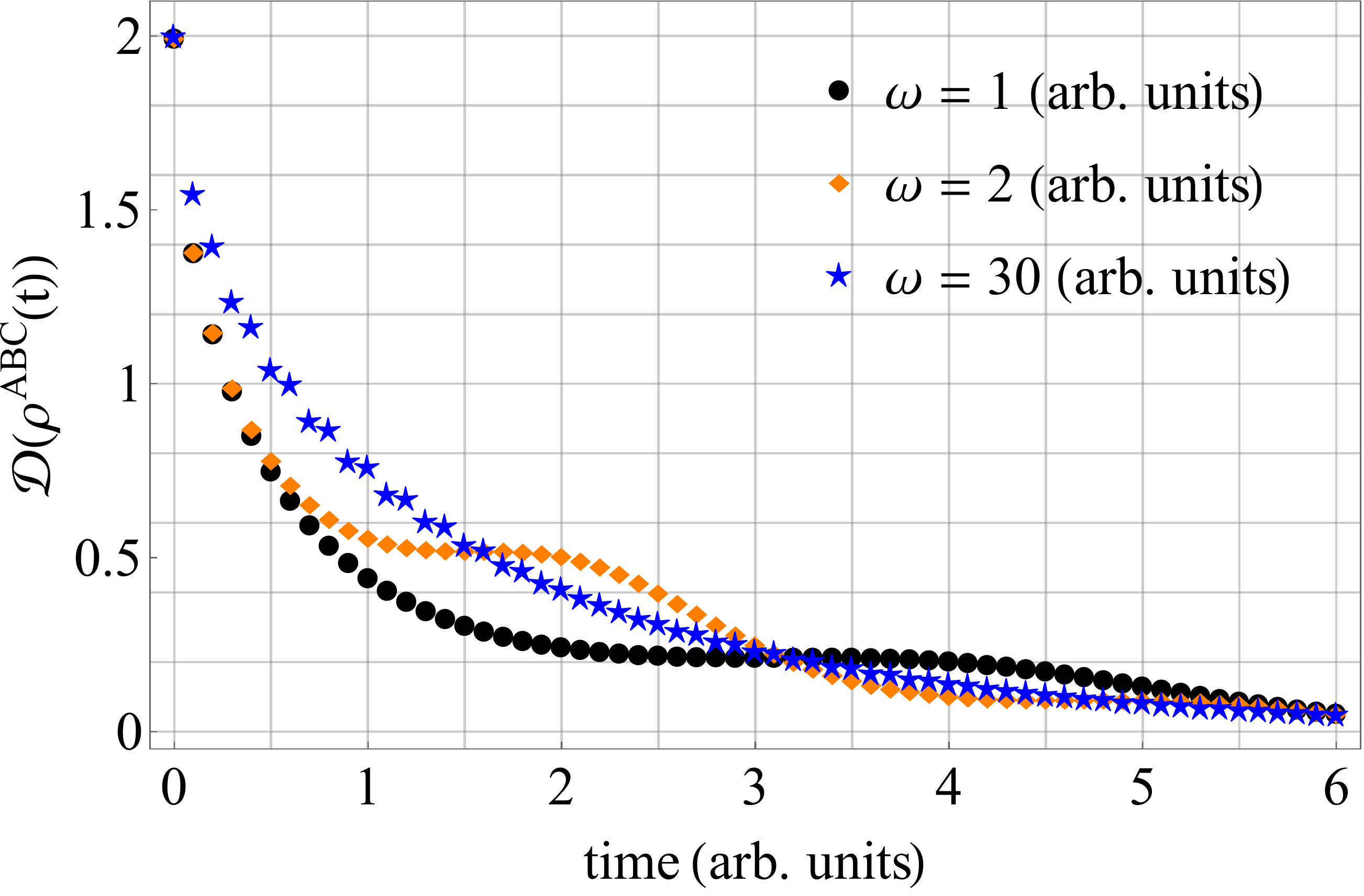}}
	\caption{Plots present the Bures distance, $\mathcal{D} (\rho^{ABC} (t))$, for three values of $\omega$. Initially, the system was represented by the W state.}
	\label{figure10}
\end{figure}

Then, in \figref{figure9}, we provide the probabilities of finding the three-qubit system in all admissible states. One observes that the probability of measuring the system in the state $\ket{111}$ grows gradually from zero towards $1$.

Finally, we can study how the entanglement declines in time by computing the Bures distance between $\rho^{ABC} (t)$ and the ultimate state $\ket{111}$, as introduced in \eqref{ex10}. In \figref{figure10}, one finds the plots of $\mathcal{D} (\rho^{ABC} (t))$ for three values of $\omega$, assuming the initial state of the system was given by \eqref{ex12}. As one can notice, the plots feature distinct shapes, which implies that the decay of entanglement is correlated with the parameters describing the dynamics.

\section{Two-qutrit entangled states}\label{twoqutrit}

To describe a qutrit evolution, we introduce a four-level cascade model that represents a physical process when the system can relax from the highest energy level $\ket{4}$ into the lower state $\ket{3}$, then into the state $\ket{2}$, and finally into the ground state denoted by $\ket{1}$. Three kinds of transition are allowable, which implies that we have three jump operators: $E_{34} := \ket{3}\!\bra{4}$, $E_{23} := \ket{2}\! \bra{3}$ and $E_{12} := \ket{1}\!\bra{2}$. We assume that the corresponding relaxation rates are given by: $\gamma_{34} (t) := e^{- \omega t} $ and $\gamma_{23} (t) = \gamma_{12} (t)= \mathrm{sin}^2 (3\, \omega t)$. This results in the generator of evolution in the matrix representation:
\begin{equation}\label{ex14}
\begin{aligned}
{}&\mathbb{L}_{4L} (t) = i \left( H_{4L}^T  \otimes \mathbb{1}_4 - \mathbb{1}_4 \otimes H_{4L} \right) + \\
&+ e^{- \omega t} \left (E_{34} \otimes E_{34} - \frac{1}{2} \mathbb{1}_4 \otimes E_{44} -\frac{1}{2} E_{44} \otimes \mathbb{1}_4  \right) + \\
& + \mathrm{sin}^2 (3\, \omega\, t) \left(E_{23} \otimes E_{23} - \frac{1}{2} \mathbb{1}_4 \otimes E_{33} - \frac{1}{2} E_{33} \otimes \mathbb{1}_4  \right)+\\
& + \mathrm{sin}^2 (3\, \omega\, t) \left(E_{12} \otimes E_{12} - \frac{1}{2} \mathbb{1}_4 \otimes E_{22} - \frac{1}{2} E_{22} \otimes \mathbb{1}_4  \right),
\end{aligned}
\end{equation}
where $H_{4L}$ denotes a four-level unperturbated Hamiltonian. The energy levels are assumed to be symmetric, i.e. $H_{4L} = \mathrm{diag} (- \mathcal{E}_2, -\mathcal{E}_1, \mathcal{E}_1, \mathcal{E}_2 )$ for $\mathcal{E}_1, \mathcal{E}_2>0$.

It can be demonstrated that the closed-form solution of a master equation with the generator \eqref{ex14} is legitimate for such initial states that do not involve the highest energy level \cite{Czerwinski2020}. Therefore, the framework of partial commutativity allows one to study the dynamics of genuine qutrit states that are spanned by the vectors $\{\ket{1}, \ket{2}, \ket{3}\}$.

To implement the framework for entangled qutrits, we introduce a two-qutrit generator
\begin{equation}\label{ex15}
\mathbb{L}_{2Q} = \mathbb{L}_{4L}^{(A)} (t) \otimes \mathbb{I}_{16}^{(B)} + \mathbb{I}_{16}^{(A)} \otimes \mathbb{L}_{4L}^{(B)} (t).
\end{equation}

Then, for any bipartite system described by an initial state $\rho^{AB} (0)$ such that neither of the subsystems involves the highest energy level, one can follow the closed-form dynamical map:
\begin{equation}\label{ex16}
\rho^{AB} (t) = \exp \left( \int_0^t \mathbb{L}_{2Q} (\tau) d \tau \right) [\rho^{AB} (0)].
\end{equation}
This method allows us to study the dynamics of two-qutrit entangled states such that each subsystem is realized as a combination of three energy levels: $\{\ket{1}, \ket{2}, \ket{3}\}$. In particular, we choose to investigate the dynamics of a maximally entangled two-qutrit state \cite{Caves2000}:
\begin{equation}\label{ex17}
\ket{\Theta}= \frac{1}{\sqrt{3}} \left(\ket{1}_A \otimes \ket{1}_B + \ket{2}_A \otimes \ket{2}_B + \ket{3}_A \otimes \ket{3}_B  \right).
\end{equation}

\begin{figure}[h!]
\centering
		\centered{\includegraphics[width=0.9\columnwidth]{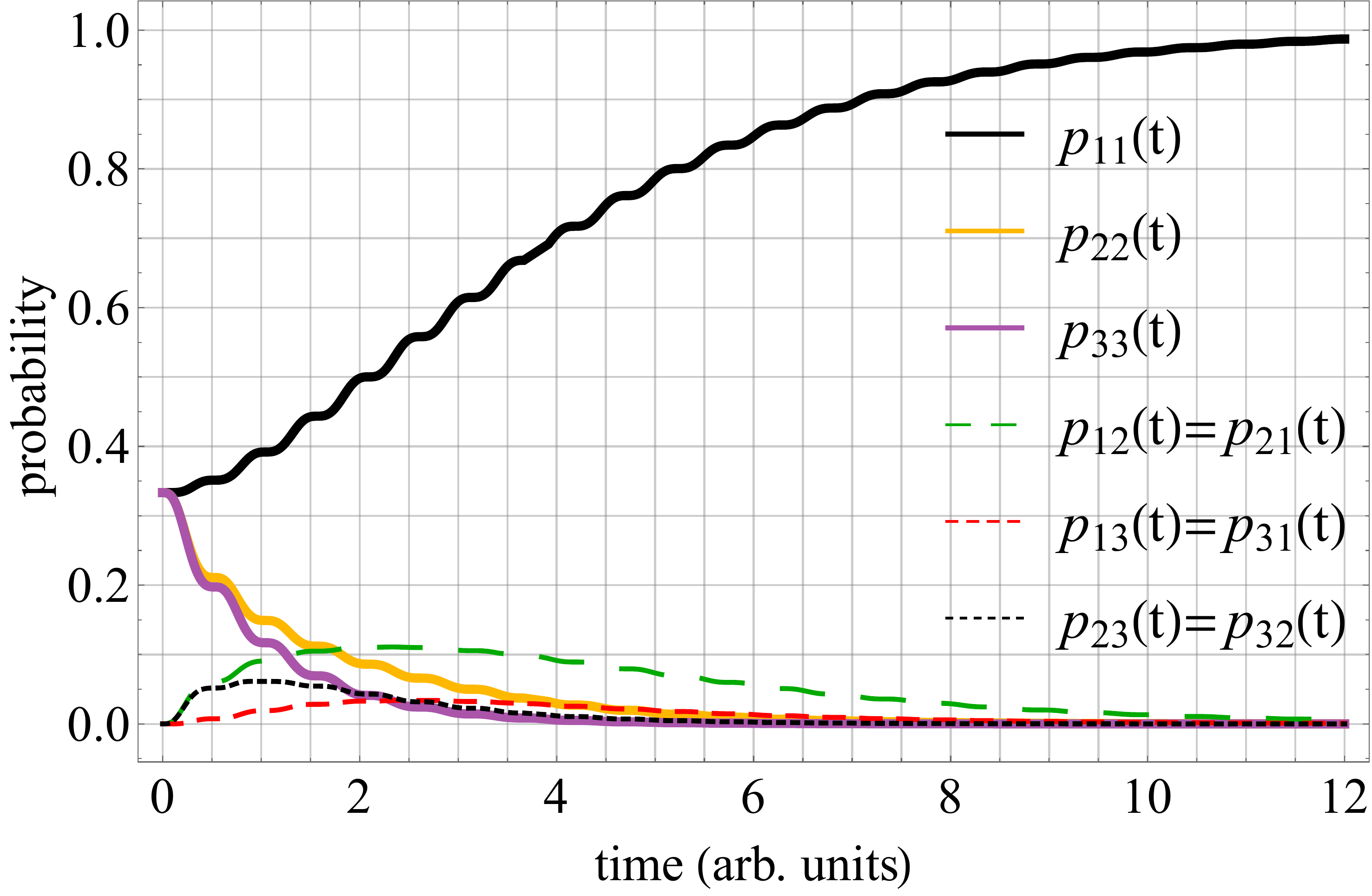}}
	\caption{Plots present the probability of finding a two-qutrit system in one of the possible states. The initial state is represented by \eqref{ex17}.}
	\label{figure11}
\end{figure}

For $\rho^{AB} (0) = \ket{\Theta}\!\bra{\Theta}$, we obtain a solution based on \eqref{ex16}, but the exact form of $\rho^{AB} (t)$ is not presented due to its complexity. Instead, in \figref{figure11}, we present the probabilities of finding the system in all of the possible configurations. As one can notice, the input state \eqref{ex17} featured perfect correlations, which means that if subsystem $A$ is found after measurement to be in a state $\ket{j}$, the subsystem $B$ is determined to be in the very same state. However, these correlations are disturbed by the dissipative generator of evolution as for $t>0$ we observe non-zero probabilities $p_{ij} (t)$ corresponding to $i\neq j$. To conclude, before the maximally entangled state \eqref{ex17} collapses into the ground level $\ket{11}$, it gets decorrelated due to the evolution governed by the time-local generator \eqref{ex15}.

\begin{figure}[h!]
\centering
		\centered{\includegraphics[width=0.9\columnwidth]{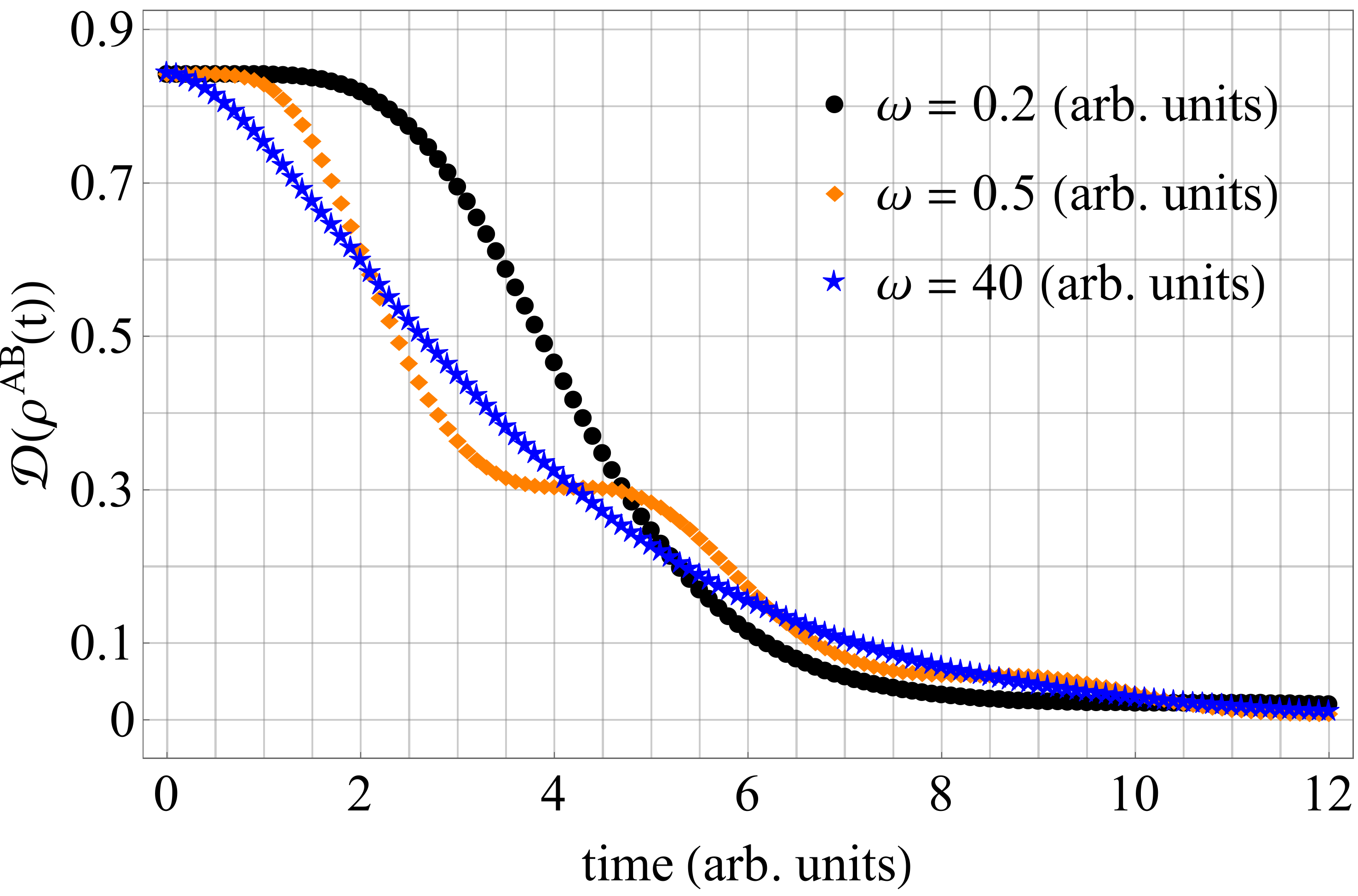}}
	\caption{Plots present the Bures distance, $\mathcal{D} (\rho^{AB} (t))$, for three values of $\omega$.}
	\label{figure12}
\end{figure}

Moreover, we investigate the dynamics of entanglement by computing the Bures distance between $\rho^{AB}  (t)$ and the final state $\ket{11}$ (we proceed analogously as in \eqref{ex10}). Again, since the state $\rho^{AB}  (t)$ approaches the separable state $\ket{11}\equiv \ket{1}_A \otimes \ket{1}_B$ with time, we consider the Bures distance as a simplified measure of entanglement. In \figref{figure12}, one finds the plots of $\mathcal{D} (\rho^{AB} (t))$ for the initial state \eqref{ex17} and three values of $\omega$. Based on the plots, one can observe that entanglement vanishes at different rates depending on the parameters characterizing the generator of evolution. Also, the functions $\mathcal{D} (\rho^{AB} (t))$ present different shapes. Such analysis allows one to track the decline of entanglement in the time domain for a given parameter.

\section{Non-Markovian evolution of two-qubit entangled states}\label{nonmarkovian}

Let us generalize the operator \eqref{tq1} by including an additional real parameter $\eta$:
\begin{equation}\label{nm1}
\begin{aligned}
{}&\mathbb{L}_{nm} (t) =  i \left( H^T  \otimes \mathbb{1}_3 - \mathbb{1}_3 \otimes H \right)  \\ & +(\mathrm{sin}^2 \omega t-\eta) \left (E_{23} \otimes E_{23} - \frac{1}{2} \mathbb{1}_3 \otimes E_{33} -\frac{1}{2} E_{33} \otimes \mathbb{1}_3  \right) \\
& +(\mathrm{cos}^2 \omega t-\eta) \left(E_{12} \otimes E_{12} - \frac{1}{2} \mathbb{1}_3 \otimes E_{22} - \frac{1}{2} E_{22} \otimes \mathbb{1}_3  \right),
\end{aligned}
\end{equation}
which implies that, depending on the value of $\eta$, the generator \eqref{nm1} may lead to: Markovian evolution, non-Markovian dynamics, or a non-physical map. Irrespective of the type of evolution, the generator \eqref{nm1} is partially commutative, which allows one to write a closed-form solution provided the highest energy state $\ket{3}$ is not included in the input state. To guarantee a physical evolution, we need to verify whether the map $\exp (\int_0^t \mathbb{L}_{nm} (\tau) d \tau) \equiv \Lambda_t$ is completely positive and trace-preserving (CPTP). Conservation of the trace is provided by the algebraic structure of the operator \eqref{nm1}, which is a time-dependent GKSL generator \cite{Gorini1976,Lindblad1976,Breuer2009}. By following the Choi's theorem on completely positive maps \cite{Choi1975}, we know that $\Lambda_t$ is CP iff $(\mathbb{I}_d \otimes \Lambda_t)[\mathcal{P}_d]\geq0$, where $\mathcal{P}_d$ denotes a projector corresponding to a maximally entangled state. In our case, $d=2$ and $\mathcal{P}_2 = \ket{\Psi^+}\!\bra{\Psi^+}$ since we reduce the dimension of the Hilbert space due to the partial commutativity constraint. However, general criteria for the map $\Lambda_t$ to be CPTP cannot be established because of the number of parameters. Therefore, we verify numerically that for $\eta=0.39$ and three selected values of $\omega$ the map $\Lambda_t$ is CPTP for all $t\geq0$.

\begin{figure}[h!]
\centering
		\centered{\includegraphics[width=0.9\columnwidth]{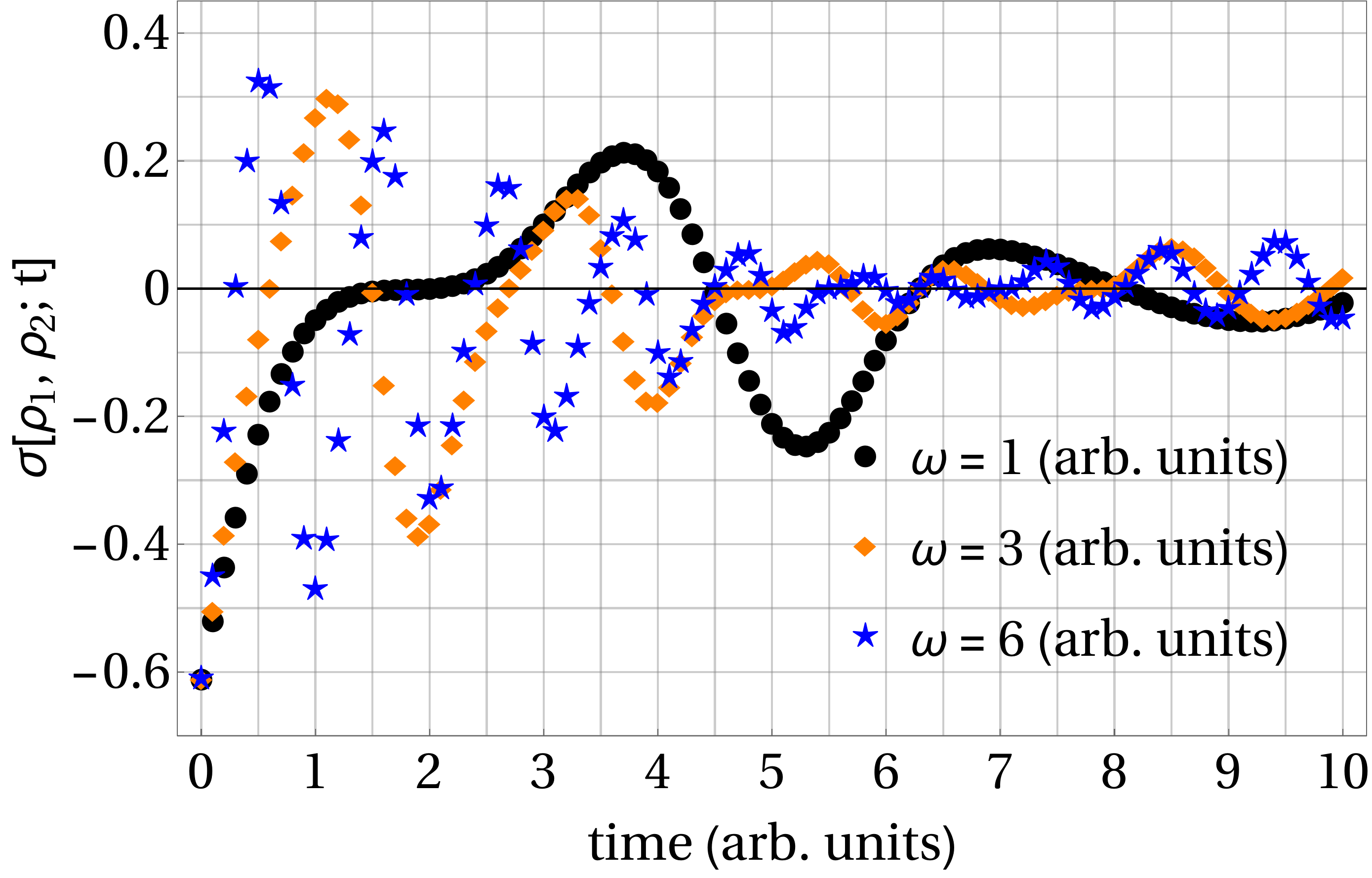}}
	\caption{Information flow, $\sigma (\rho_1, \rho_2;t)$, for three values of $\omega$.}
	\label{sigma}
\end{figure}

Non-Markovian behavior of the map $\Lambda_t$ can be demonstrated by following the criterion given by Breuer, Lane, and Piilo (henceforth: BLP criterion) \cite{Breuer2009}. They constructed a general measure for the degree of non-Markovianity in open quantum systems. According to the BLP criterion, a dynamical map $\Lambda_t$  is Markovian iff
\begin{equation}\label{nonMarkovian}
\sigma (\rho_1, \rho_2;t) := \frac{d}{d \,t} \| \Lambda_t (\rho_1 - \rho_2)\| \leq 0
\end{equation}
for all pairs of input states $\rho_1$ and  $\rho_2$. In \eqref{nonMarkovian}, we use $\| .\|$ to denote the trace norm, i.e., $\|X\| = \tr \sqrt{X X^{\dagger}}$. The BLP criterion can be implemented to demonstrate non-Markovian effects. The figure $\sigma (\rho_1, \rho_2;t)$ can be interpreted as an information flow and, as a results, $\sigma (\rho_1, \rho_2;t) \leq 0$ implies that the information is lost over time. On the other hand, $\sigma (\rho_1, \rho_2;t)>0$ indicates a backflow of information from environment to the system, which is a proof of non-Markovian effects. In our application, we select $\rho_1 = \ket{\Phi^+}\!\bra{\Phi^+}$ and $\rho_2 = 1/2 \: \ket{\Phi^+}\!\bra{\Phi^+} + 1/2 \: \mathbb{I}_4$. To demonstrate that our map $\Lambda_t$ features non-Markovianity, we plot $\sigma (\rho_1, \rho_2;t)$, for three exemplary values of $\omega$. The results are presented in \figref{sigma}. One can observe that we have both positive and negative values of $\sigma (\rho_1, \rho_2;t)$, which means that during evolution information can flow in both directions (from system to the environment and vice versa).

Then, we study the joint dynamics of a two-qubit system: $\rho^{AB} (t) = (\Lambda_t \otimes \Lambda_t) \rho^{AB} (0)$, where as the input state we take $\rho^{AB} (0)= \ket{\Phi^+}\!\bra{\Phi^+}$. 

\begin{figure}[h!]
\centering
		\centered{\includegraphics[width=0.9\columnwidth]{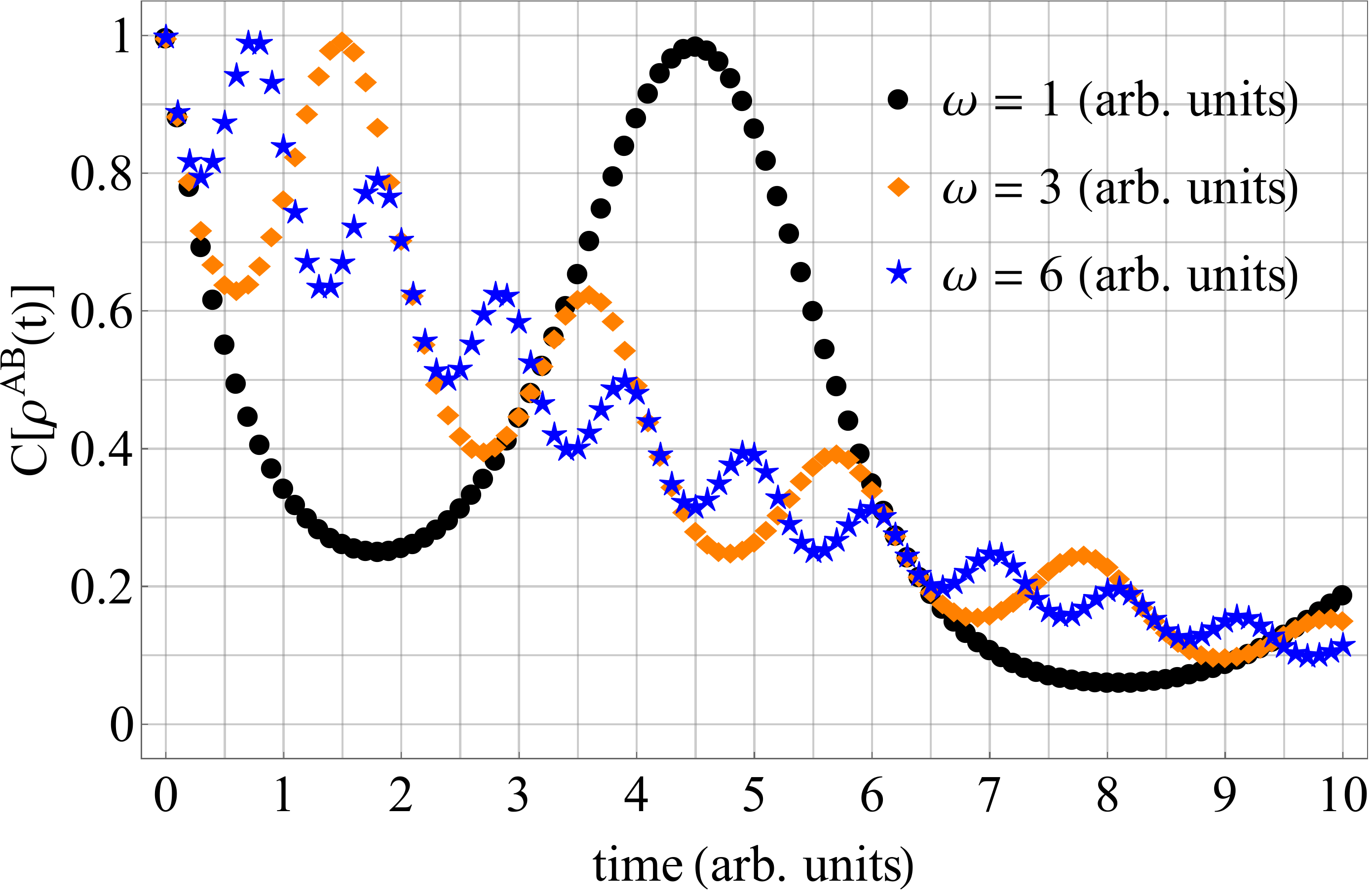}}
	\caption{Concurrence, $C[\rho^{AB} (t)]$, for three values of $\omega$ and the initial state: $\rho^{AB} (0)= \ket{\Phi^+}\!\bra{\Phi^+}$.}
	\label{figure13}
\end{figure}

\begin{figure}[h!]
\centering
		\centered{\includegraphics[width=0.9\columnwidth]{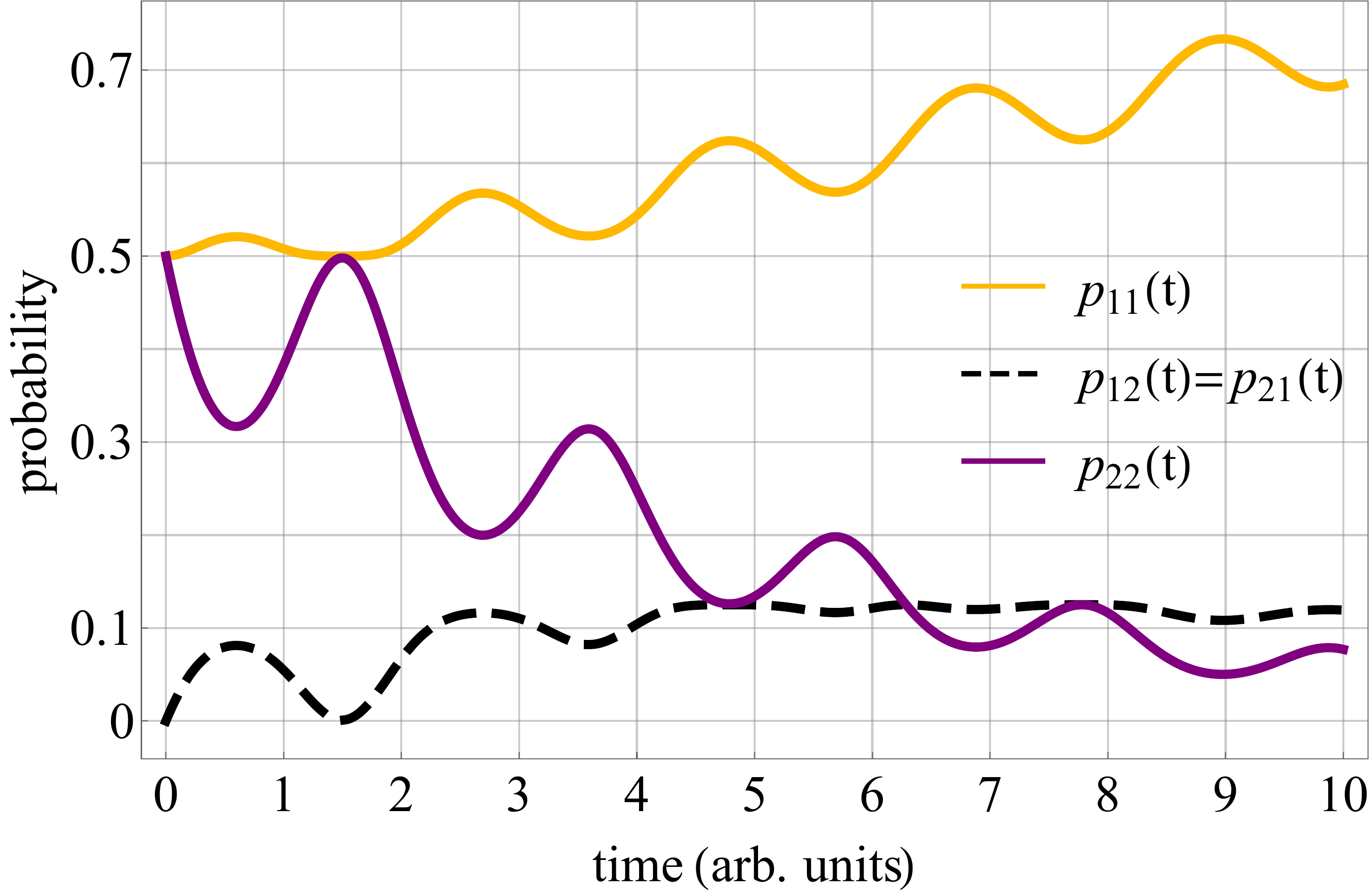}}
	\caption{Plots present the probability of finding the system $\rho^{AB} (t)$ in one of the possible states, assuming $\omega = 2$ (arb. units).}
	\label{figure14}
\end{figure}

In \figref{figure13}, one can observe the concurrence versus time for three specific values of $\omega$ (the same as those used to depict \figref{sigma}). The plots feature clear non-Markovian effects. We start from a maximally entangled state and, initially, the concurrence decreases. However, we observe a backflow of information from the environment to the system during the dynamics. After the first decline, the entanglement is restored by non-Markovianity, and the concurrence again reaches its maximum value. Then, the concurrence oscillates, and the local maxima can be attributed to non-Markovianity.

The backflow of information caused by non-Markovianity can also be observed by studying the probabilities corresponding to finding the system (upon measurement) in an admissible state. In \figref{figure14}, we present plots for a selected value of $\omega$. In particular, the first backflow is evident, when the probabilities return to their initial values, and the entanglement is regained. Later on, the probabilities oscillate.

\section{Discussion and outlook}

In the article, we implemented the framework of partial commutativity to study the dynamics of entangled states governed by time-dependent generators. The method allows one to obtain a closed-form solution of a master quantum equation for a subset of initial states determined by the Fedorov theorem. Consequently, one can investigate the dynamics of lower-dimensional subsystems immersed in the original Hilbert space. In particular, the framework proved to be an efficient tool for entanglement analysis. In this paper, we investigated two-qubit and three-qubit entangled states, as well as the evolution of entangled qutrits. In each case, the framework enabled us to study in detail the dynamics of celebrated types of entanglement, which demonstrates how nonunitary forms of decoherence affect entanglement.

Entangled states are considered a key resource in quantum computation and communication. Therefore, one would like to preserve a sufficient degree of entanglement for the longest achievable period of time. On the other hand, the theory of open quantum systems indicates that interactions between the system and its environment can lead to a decrease in the amount of entanglement. Therefore, it appears relevant to study the impact of different evolution models on the amount of entanglement. The framework of partial commutativity allows one to follow the decay of entanglement driven by time-local dissipative generators with positive relaxation rates.

For negative decoherence rates, the framework allows one to witness non-Markovian effects. In the Markovian regime, there is a continuous flow of information from the system to the environment. However, if we go beyond this approximation, one can observe a backflow of information, which leads to an increase of the concurrence during the evolution and, as a result, entanglement can be restored. These findings are in accordance with other studies devoted to non-Markovian effects on the dynamics of entanglement \cite{Bellomo2007}. Howbeit, in the present work, we demonstrated that the maximum degree of entanglement could be regained due to the dynamics governed by a time-local generator.

Recent advances in experimental techniques and fabrication of quantum materials have led us to circumstances where non-Markovian effects became crucial, opening new arenas for scientific exploration. Non-Markovian dynamics of open quantum systems is often studied within the memory kernel approach \cite{Vega2017}, which utilizes the Nakajima-Zwanzig equation \cite{nakajima,zwanzig}. However, the present paper indicates that partial commutativity can also be an efficient tool for examining the properties of non-Markovian evolution emerging from time-local quantum generators. In particular, the framework can be implemented to transfer an input state to the target state strictly along the designed trajectory, including non-Markovian reservoir, cf. \cite{Wu2022}.

The framework described in this article is versatile and can be implemented to other multilevel systems. For a given quantum generator, the condition of partial commutativity allows one to cut out a subset of initial quantum states for which the dynamical map can be written in the closed form. This aspect is connected with reducing the dimension, which implies that certain levels have to be dropped for the closed-form solution to be legitimate. The subsets of allowable states may have different structures, depending on the algebraic properties of the generator of evolution as well as on specific values of the parameters characterizing dynamics.

In the future, the framework will be developed to investigate dynamics governed by other classes of time-dependent generators, including non-Markovian evolution. The goal of further research is to go beyond dissipative generators that describe the process of relaxation. The framework is expected to provide significant insight into atom-photon interactions. The ability to control quantum dynamics in such processes as laser cooling can contribute to the advancement in quantum computing with single atoms.

\section*{Acknowledgement}

The research was supported by the National Science Centre in Poland, grant No. 2020/39/I/ST2/02922.

\end{document}